\documentclass[journal=jctcce,manuscript=article]{achemso}
\usepackage{amsmath, amsfonts}
\usepackage{booktabs}
\usepackage[dvipsnames]{xcolor}
\usepackage{hyperref}
\usepackage{amssymb}
\usepackage{amsmath}
\usepackage{bbold}
\usepackage{color}
\usepackage{bm}
\usepackage{epsfig}
\usepackage[caption=false]{subfig}
\usepackage{float}
\usepackage{subeqnarray}
\usepackage{array}
\usepackage{booktabs}
\usepackage[version=4]{mhchem}
\usepackage{tikz}
\usetikzlibrary{quantikz2}

\title{Simulating Non-Markovian Quantum Dynamics on NISQ Computers Using the Hierarchical Equations of Motion}
\author{Xiaohan Dan}
\affiliation{Department of Chemistry, Yale University, New Haven, CT 06520,
USA}
\author{Eitan Geva}
\affiliation{Department of Chemistry, University of Michigan, Ann Arbor, MI 48109, USA}
\author{Victor S. Batista}
\affiliation{Department of Chemistry, Yale University, New Haven, CT 06520, USA}
\alsoaffiliation{Yale Quantum Institute, Yale University, New Haven, CT 06511, USA}
\email{victor.batista@yale.edu}

\begin{document}

\begin{abstract}
Quantum computing offers promising new avenues for tackling the long-standing challenge of simulating the quantum dynamics of complex chemical systems, particularly open quantum systems coupled to external baths. However, simulating such non-unitary dynamics on quantum computers is challenging 
since quantum circuits are specifically designed to carry out unitary transformations. 
Furthermore, chemical systems are often strongly coupled to the surrounding environment, rendering the dynamics non-Markovian and beyond the scope of Markovian quantum master equations like Lindblad or Redfield.
In this work, we introduce a quantum algorithm designed to simulate non-Markovian dynamics of open quantum systems. 
Our approach enables the implementation of arbitrary quantum master equations on noisy intermediate-scale quantum (NISQ) computers. We illustrate the method as applied in conjunction with the numerically exact hierarchical equations of motion (HEOM) method. The effectiveness of the resulting quantum HEOM algorithm is demonstrated as applied to simulations of the non-Lindbladian electronic energy and charge transfer dynamics in models of the carotenoid-porphyrin-\ce{C60} molecular triad dissolved in tetrahydrofuran and the Fenna-Matthews-Olson complex. 
\end{abstract}
\maketitle

\section{Introduction}
Simulations of open quantum systems are essential for theoretical studies of a wide range of chemical processes, including charge transfer, energy transfer, and proton transfer in solutions or biological systems.~\cite{breuer02,nitzan06,weiss12}
However, accurate simulations of such systems are challenging and often rely on the Markovian approximation. Here, we introduce a quantum algorithm for simulations of non-Markovian open quantum systems on noisy intermediate-scale quantum (NISQ) computers.

Traditionally, dynamical simulations of open quantum systems relied on Markovian quantum master equations of the Redfield or Lindblad types. These approaches assume that the system is weakly coupled to a bath so that the coupling can be treated within second-order perturbation theory.~\cite{wangsness53,redfield57,pollard96,nitzan06,carmichael13,manzano20,alicki87} While effective in numerous scenarios, these methods fall short when chemical systems exhibit non-Markovian behavior. This occurs under conditions such as strong system-environment coupling, low temperatures, structured or finite reservoirs, or initial correlations between the system and its environment.~\cite{zheng13,breuer16,vega17,dan22,dan23}

To address non-Markovian dynamics, various non-perturbative methods have been developed, including tensor train thermo-field dynamics (TT-TFD),~\cite{gelin17,borrelli21b,lyu23,lyu24} the hierarchical equations of motion (HEOM),~\cite{tanimura89,tanimura06,Tanimura20,shi09c,shi18,dan23,dan23b}
the multilayer multiconfiguration time-dependent Hartree (ML-MCTDH),~\cite{wang03,wang15b} path integral techniques,~\cite{topaler93,makri95,weiss08b,segal10,weiss13,makri20,bose22,walters24,seneviratne24}
and the generalized quantum master equation (GQME).~\cite{zwanzig61a,shi03g,shi04a,zhang06,kelly16,montoya16,montoya17,pfalzgraff19,mulvihill19,mulvihill21b,mulvihill22,ng21,dan22,lyu23}

Recent advancements in quantum computing hardware have opened new possibilities for 
simulating quantum dynamics on quantum computers.\cite{nielsen10,kassal11,georgescu14,cao19,ollitrault21,delgado24}
Most quantum algorithms have focused on closed systems, where unitary dynamics can be directly mapped onto quantum circuits.\cite{wiebe11,ollitrault21,yao21}
However, the inherently non-unitary nature of open quantum systems dynamics presents a unique challenge for quantum computing, as quantum circuits are specifically designed to implement unitary transformations.\cite{nielsen10,sweke15,hu20}

Efforts to bridge this gap and map non-unitary dynamics into a unitary framework have led to the development of the linear combination of unitaries (LCU),\cite{schlimgen21,schlimgen22b,li24} 
dilation methods,\cite{sweke15,hu20,head-marsden21,schlimgen22,wang23} 
quantum imaginary time evolution,\cite{mcardle19,motta20,kamakari22} 
density-matrix purification,\cite{schlimgen22c}
and variational quantum algorithms.~\cite{li17b,chen20,endo20,lee22} 
However, most quantum algorithms are tailored to systems governed by the Lindblad quantum master equation, which assumes weak system-bath coupling and Markovian dynamics. While these algorithms have been applied to simulate relatively simple models, such as spontaneous emission and two-level systems,~\cite{hu20,schlimgen21,schlimgen22b,schlimgen22,kamakari22,bacon01,cleve16,luo24} their application to more complex model systems like the transverse field Ising model,\cite{schlimgen22b,kamakari22,endo20,luo24}
and the Fenna-Matthews-Olson (FMO) complex,\cite{hu22}
often relies on ad-hoc choices of Lindblad operators, which may not fully capture dephasing or damping processes induced by the environment.

In this work, we implement the numerically exact HEOM approach with the dilation method to simulate non-Markovian open quantum system dynamics on quantum computers. In what follows, we will refer to this implementation as qHEOM.
We illustrate the capabilities of the qHEOM algorithm on IBM quantum computers as applied to simulations of charge transfer dynamics in a solvated molecular triad \cite{tong20} and electronic energy transfer dynamics in the FMO complex. 
Additionally, we evaluate the applicability of the time-convolutionless (TCL) Redfield equation with the rotating wave approximation (RWA), which serves as the quantum master equation in the Lindblad form for the model systems under consideration.
By comparing the results obtained by integrating the TCL-Redfield equation with the RWA with those obtained from the qHEOM method, we illustrate the limitations of the TCL-Redfield equation across the parameter regimes relevant to electronic charge and energy transfer processes.

Our qHEOM method belongs to the family of recently proposed quantum algorithms for non-Markovian evolution.~\cite{sweke16,head-marsden21,wang23,walters24,seneviratne24,li24} Using the Sz.-Nagy dilation theorem,~\cite{levy14} Wang \emph{et al.} \cite{wang23} dilated the propagator obtained from the GQME.
Walters \emph{et al.} \cite{walters24} constructed the propagator time series that spans a memory time using the path integral approach, and dilated it into unitary gates. Seneviratne \emph{et al.} \cite{seneviratne24} used dilation based on singular value decomposition\cite{schlimgen22} (SVD dilation) to dilate the Kraus operators that are calculated through path integral.
Li \emph{et al.}\cite{li24} presented a quantum algorithm based on the linear combination of unitaries (LCU) approach~\cite{schlimgen22b}
to implement the numerically exact dissipaton-embedded quantum master equation in second quantization (DQME-SQ). 

The qHEOM algorithm offers several key advantages over the methods mentioned above. First, it employs the SVD dilation methodology that can be applied to dilate propagators from a wide range of master equations. This SVD approach essentially decomposes the propagator into a sum of two unitary operators that require much fewer shots than the traditional LCU method based on the Taylor expansion.~\cite{schlimgen22b} Second, when compared to Sz.-Nagy dilation, the SVD approach dilates only the diagonal matrix of singular values, significantly reducing the circuit depth. Additionally, the diagonal unitary operator of singular values can be efficiently implemented using the Walsh operator representation,~\cite{welch14} further reducing circuit complexity. Another important advantage of qHEOM is that it employs projection operators to map vectors from the HEOM space to the state vector for quantum computing. The flexibility in choosing the projection subspace allows us to select a smaller subspace and reduce the dimension of the propagator, thereby decreasing the number of qubits and circuit complexity on NISQ devices. 

The outline of this paper is as follows. Section \ref{Sec: qHEOM_algorithm} introduces our quantum algorithm for the propagation of non-Markovian dynamics based on SVD dilation. Section \ref{Sec: num_methods} describes the simulation methods based on the HEOM and the TCL-Redfield equation with the RWA. Section \ref{Sec: ModelSystem} presents the model systems used for electronic charge and energy transfer simulations. Section \ref{Sec: result} compares the simulation results from HEOM and the TCL-Redfield equation with the RWA on classical computers, along with qHEOM simulations. Section \ref{Sec: conclusion} concludes the paper with a summary of findings and future directions.
Overall, our work demonstrates the potential of a novel quantum computing algorithm to simulate complex non-Markovian dynamics, providing insights into quantum phenomena in chemical systems beyond the limitations of approximate methods. 

\section{Quantum Algorithm for Open Quantum Systems}\label{Sec: qHEOM_algorithm}

\subsection{Time Evolution of the Register State Vector}\label{Sec: qb_state_evolve}

The state vector $|\Phi\rangle$ represents the state of the qubits that make up the register of the quantum circuit. Its time evolution is mapped to the evolution of the reduced density matrix elements that describe the non-Markovian dissipative dynamics of the system of interest. To reduce the depth of the circuits, we employ projection operators that allow us to propagate subsets of the reduced density matrix elements. This approach is exact and reduces both the number of gates and the number of qubits required for simulation, allowing for parallel quantum computing of subsets of matrix elements without introducing any approximation. 

The procedures for projecting the initialized density matrix elements and encoding them into the state vector $|\Phi(0) \rangle$ are described in section~\ref{sec:proj_propagator}. The time-evolved state vector $|\Phi(t) \rangle$ is obtained as
\begin{equation}\label{Eq:qProp}
    |\Phi(t)\rangle = G(t) |\Phi(0)\rangle \;,
\end{equation}
where $G(t)$ represents the non-unitary propagator corresponding to the numerical method of choice. This non-unitary evolution reflects the open nature of the system interacting with its surrounding environment. 
As described in section~\ref{sec:lcu}, $G(t)$ is implemented as a linear combination of unitaries. 

The overall accuracy of the simulation depends on the accuracy of the propagator of choice. Here, we encode the propagator of HEOM which yields numerically exact dynamics, incorporating non-Markovian effects and enabling precise simulations only limited by the number of shots and the level of noise in the quantum device.

\subsection{Turning the Propagator into Linear Combination of Unitaries}
\label{sec:lcu}

We decompose the non-unitary propagator into a linear combination of unitaries implementing the singular value decomposition (SVD) method proposed by Schlimgen \emph{et al.}~\cite{schlimgen22} 
The procedure starts with the SVD of the propagator $G(t)$:
\begin{equation}
\label{eq:svd}
{G}(t) = U \Sigma {V}^\dag \;,
\end{equation}
where $U$ and $V$ are unitary matrices and $\Sigma$ is a diagonal matrix containing the singular values. The singular value matrix $\Sigma$ is expressed as a linear combination of two diagonal unitary matrices ${\Sigma}_+$ and ${\Sigma}_-$, 
\begin{equation}\label{Eq:singval_Gt}
\Sigma = \frac{\sigma_0}{2}({\Sigma}_+ + {\Sigma}_-) \;,
\end{equation}
where the diagonal elements of ${\Sigma}_+$ and ${\Sigma}_-$ are defined as
\begin{equation}
\label{eq:sumf}
({\Sigma}_{\pm})_{jj} = \tilde{\sigma}_{j} \pm i \sqrt{{1-\tilde{\sigma}_{j}^2}} \;,
\end{equation}
with $\tilde{\sigma}_{j} = \sigma_{j}/\sigma_{0}$. Here, $\sigma_{j}$ is the $j$-th singular value of ${G}(t)$ and $\sigma_0$ is the largest singular value. Using ${\Sigma}_+$ and ${\Sigma}_-$, the propagator ${G}(t)$ is decomposed into the linear combination of two unitaries, as follows: 
\begin{equation}
\label{eq:lcu}
\begin{split}
{G}(t) &= \frac{\sigma_0}{2} 
({U}{\Sigma}_+ {V}^\dag+ {U}{\Sigma}_- {V}^\dag) \;\;.
\end{split} 
\end{equation}

\subsection{Quantum Circuit for Non-Unitary Propagation}

\begin{figure}[h]
\begin{quantikz}
\lstick{$\ket{\Phi(0)}$} &\gate{{V}^\dag} & \gate[2]{{U}_\Sigma} &\gate{{U}} & \meter{} \\
\lstick{$\ket{0}$} & \gate{H}   & & \gate{H} &\meter{} 
\end{quantikz}
\caption{Quantum circuit for SVD dilation,\cite{schlimgen22,seneviratne24} with $H$ the single qubit Hadamard gate, while $V$ and $U$ are defined according to eq~\ref{eq:svd}.}
\label{Fig:qc_dilation}
\vspace{1em}
\end{figure}
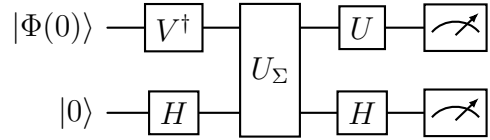

The linear combination of two unitaries, introduced by eq~\ref{eq:lcu}, can be readily implemented by a quantum circuit with a one-qubit dilation, using unitary gates as shown in Fig.~\ref{Fig:qc_dilation}. 
To achieve this, we define a diagonal unitary operator ${U}_{\Sigma} = \Sigma_+ \oplus \Sigma_-$ that acts on both the main and ancilla qubits, as follows: 
\begin{align}\label{Eq:dilat_sigma}
{U}_{\Sigma} &= \left( 
\begin{array}{cc}
{\Sigma}_+ & 0	 \\
0 & {\Sigma}_-   \;\;
\end{array} \right) \;.
\end{align} 
The output of the quantum circuit is the following state: 
\begin{equation}\label{Eq:state_qc_svd}
\begin{aligned}
\frac{1}{2} \left(
\begin{array}{c}
{U}({\Sigma}_+ + {\Sigma}_-){V}^\dag
|\Phi(0) \rangle	 \\
{U}({\Sigma}_+ - {\Sigma}_-){V}^\dag
|\Phi(0) \rangle \;\;
\end{array} \right) 
= \frac{1}{\sigma_0} \left(
\begin{array}{c}
G(t)|\Phi(0) \rangle	 \\
|\phi\rangle  \;\;
\end{array} \right) \;. 
\end{aligned}
\end{equation}
Therefore, when the ancilla is in state $|0\rangle$, we obtain the desired state $G(t)|\Phi(0) \rangle/\sigma_0 = {U}({\Sigma}_+ + {\Sigma}_-){V}^\dag
|\Phi(0) \rangle/2$ (up to the normalization factor $\sigma_0$). When the ancilla is in state $|1\rangle$,  we obtain the state $|\phi\rangle = {U}({\Sigma}_+ - {\Sigma}_-){V}^\dag
|\Phi(0) \rangle/2$, which is discarded.

\subsection{Efficient Implementation of Diagonal Unitary Operators Using Walsh Operators}
The diagonal unitary ${U}_{\Sigma}$, introduced by eq~\ref{Eq:dilat_sigma}, entangles the main and ancilla qubits. Upon compilation, this operation typically generates deep circuits which represent the computational bottleneck for computations on NISQ devices. 
However, efficient implementations of diagonal unitaries can be achieved using the Walsh operator representation.~\cite{welch14} For example, Seneviratne {\em et al.} have implemented Walsh operators to compile the singular value diagonal matrix of Kraus operators.~\cite{seneviratne24} Here, we apply this technique to optimize the implementation of ${U}_{\Sigma}$.

$N$-dimensional ($N=2^n$) diagonal unitary matrices ${U}_{\Sigma}$, can be expressed, as follows: 
\begin{equation}\label{Eq:U_eif}
{U}_{\Sigma} = e^{i \hat{F}},
\end{equation}
where $\hat{F}$ is a real diagonal matrix. Its diagonal elements $f_k$ define the diagonal elements of $U_{\Sigma}$: $(U_{\Sigma})_{kk} = e^{i f_k}$. When ${U}_{\Sigma}$ is defined according to eq~\ref{Eq:dilat_sigma}, with matrix elements defined
according to eq~\ref{eq:sumf}, we have $f_k = \text{arccos}~\tilde{\sigma}_{k}$, for $k=0,1,...,N/2-1$, and $f_k = -\text{arccos}~\tilde{\sigma}_{k-N/2}$, for $k=N/2,...,N-1$. 

The real diagonal matrix $\hat{F}$ can be represented in the $n$-qubit Pauli-Z and identity basis through 
\begin{equation}\label{Eq:F_in_Pauli}
   \hat{F} = \sum_{j=0}^{N-1} a_j \hat{\rm w}_j \,,
\end{equation}
where $a_j$, are the Walsh coefficients, and $\hat{\rm w}_j$, are the Walsh operators ({\em i.e.}, tensor products of the one-qubit identity and Pauli-Z matrices):
\begin{equation}
    \hat{\rm w}_j = (\hat{Z}_1)^{j_1} \otimes  (\hat{Z}_2)^{j_2} \otimes \cdots \otimes (\hat{Z}_n)^{j_n} \;,
\end{equation}
where $(\hat{Z}_l)^0 \equiv \hat{I}$ (identity gate) and $(\hat{Z}_l)^1 \equiv \hat{Z}$ (Pauli-Z gate), act on the $l$-th qubit. Here, "$\otimes$" is the Kronecker product, while $j_l \in \{0, 1\}$ are the bits of the binary expansion of $j$ ($j=\sum_{l=1}^n j_l 2^{l-1}$). 
The elements of the diagonal matrix $\hat{\rm w}_j$ are defined as
\begin{equation}
\begin{split}
    [\hat{\rm w}_j]_{kk} &= \langle k| \hat{\rm w}_j|k\rangle,\\
    &= \langle k_1 k_2 \cdots k_n| \hat{\rm w}_j \ket{k_1 k_2\cdots k_n},\\
    &= \prod_{l=1}^n \langle k_l| (\hat{Z}_l)^{j_l} |k_l\rangle,\\
    &= (-1)^{\sum_l j_l k_l} \,,
\end{split}
\end{equation}
where $k_l \in \{0, 1\}$ are the bits in the binary expansion of $k$, with $k = \sum_{l=1}^{n} k_l~2^{(n-l)}$, while $\ket{k_1 k_2\cdots k_n} = \ket{k_1} \otimes \ket{k_2} \otimes \cdots \otimes \ket{k_n}$ with $\ket{0} = [1,0]^T$ and $\ket{1} = [0,1]^T$.

The Walsh coefficients $a_j$ are obtained from the Hilbert-Schmidt inner products of $\hat{F}$ and the Walsh operators $\hat{\rm w}_j$, as follows: 
\begin{equation}
\label{eq:walsh_coef}
\begin{split}
    a_j &= \frac{1}{N} {\rm Tr} \left[ \hat{\rm w}_j \,\hat{F} \right],\\
    &= \frac{1}{N} \sum_{k=0}^{N-1} [\hat{\rm w}_j]_{kk} \, f_k,\\ 
    &= \frac{1}{N} \sum_{k=0}^{N-1} (-1)^{\sum_l j_l k_l} f_k \,.
\end{split}
\end{equation}
This transformation between $a_j$ and $f_k$, introduced by eq~\ref{eq:walsh_coef}, is the so-called Walsh–Fourier transform.~\cite{walsh1923,schipp90} 

Having the Walsh representation of $\hat{F}$, introduced by eq~\ref{Eq:F_in_Pauli}, and noting that Walsh operators commute ({\em i.e.}, $[\hat{\rm w}_j,\hat{\rm w}_k] = 0$), the ${U}_{\Sigma}$ can be written as 
\begin{equation}
    {U}_{\Sigma}  = \prod_{j=0}^{N-1} e^{i a_j \hat{\rm w}_j} \,.
\end{equation}
Each gate $e^{i a_j \hat{\rm w}_j}$ can be readily implemented by using CNOT and Z-rotation gates, as described in Ref.~\citenum{welch14}. The circuit can be further optimized by recognizing that the gates $e^{i a_j \hat{\rm w}_j}$ commute. By rearranging the indices $j$ using the Gray code and leveraging commutation properties of CNOT gates, we can reduce the number of CNOT gates, as implemented in Ref.~\citenum{welch14}. 

\section{Methods for Open Quantum System Dynamics}\label{Sec: num_methods}

\subsection{Model Hamiltonian}

We simulate the dynamics of quantum systems coupled to harmonic baths, as described by the Hamiltonian 
\begin{equation}\label{Eq:Htot}
{H}_T = {H}_S + {H}_B + {H}_I \;,
\end{equation}
where ${H}_S$ is the Hamiltonian of the  system, ${H}_B$ is the bath Hamiltonian, and ${H}_I$ 
describes the system-bath interaction. 
We assume that the bath consists of harmonic modes and that the system-bath coupling is linear in the coordinates of these modes, such that: 
\begin{equation}\label{Eq:bath_coup_general}
\begin{split}
    H_B &= \sum_{mj} \frac{p_{mj}^2}{2M_{mj}} + \frac{1}{2} M_{mj} \omega_{mj}^2 x_{mj}^2 \,,\\
    {H}_I &= \sum_m A_m \otimes B_m \equiv -\sum_m A_m \otimes \sum_j c_{mj}x_{mj} \;.
\end{split}
\end{equation}
Here, to maintain generality, we consider multiple distinct harmonic baths indexed by $m$. Each bath contains multiple modes, where $x_{mj}$, $p_{mj}$, $M_{mj}$, and $\omega_{mj}$ are the coordinate, momentum, mass, and frequency of the $j$-th mode in the $m$-th bath. The Hermitian system operator $A_m$ couples to the collective coordinate $B_m =-\sum_j c_{mj}x_{mj}$ of the $m$-th bath, with the coupling strength between the system and the $mj$-th bath mode being given by $c_{mj}$.

For the aforementioned system-bath model, the influence of the environment on the system can be fully characterized by the reservoir correlation function: \cite{breuer02,ishizaki09a,weiss12} 
\begin{equation}\label{Eq:Ct_jome}
C(t) = \frac{1}{\pi}\int_0^\infty d\omega \: J(\omega) [\coth(\frac{\beta\omega}{2})\cos(\omega t)-i\sin(\omega t)] \;,
\end{equation}
where $\beta = (k_BT)^{-1}$ is the inverse temperature, and $J(\omega)$ is the spectral density, defined as 
\begin{equation}
J(\omega) = \frac{\pi}{2}\sum_j \frac{c_{j}^2}{M_{j} \omega_{j}} \delta(\omega-\omega_{j}) \;\;.
\end{equation}
For simplicity, we consider the environment to be identical for each $m$ ($c_{mj}\equiv c_j$), thus omitting the dependence on $m$ for $C(t)$ and $J(\omega)$.
Finally, we assume the spectral density has a Debye form: 
\begin{equation}\label{Eq:Debye}
J(\omega)=\frac{\eta\omega\omega_c}{\omega^2+\omega_c^2}  \;,
\end{equation}
where $\eta$ is the coupling strength, and $\omega_c$ characterizes the width of the spectral density.

\subsection{Hierarchical Equations of Motion}
\label{sec:secheom}
The HEOM approach decomposes the reservoir correlation function 
into a sum of exponentials:  
\begin{equation}\label{Eq:Ct_SOP}
C (t) = \sum_{k} d_{k} e^{-v_{k} t} \;\;,
\end{equation}
where $v_k$ and $d_k$ are the frequencies and coefficients of the effective modes, respectively. For the Debye spectral density, $v_k$ and $d_k$ are analytically given by:\cite{shi09c,neill23}
\begin{align}
    v_1 &= \omega_c \;,\\
    v_k &\equiv \frac{2\pi (k-1)}{\beta} \;;\quad k>1 ,
\end{align}
and
\begin{align}
    d_1 &= \frac{\eta \omega_c }{2}[\cot(\beta\omega_c/2)-i] \;,\\
    d_k &= \frac{2}{\beta} \frac{\eta v_k \omega_c}{v_k^2-\omega_c^2}\;;\quad k>1 .
\end{align}
This decomposition transforms the original model of a system coupled to infinite bath modes into a model of a system interacting with a finite number of effective modes.\cite{ke22,wang22,li24}

In HEOM, the density matrices can be combined into a single state vector in the tensor product Hilbert space of the system and effective modes, as follows:
\begin{equation}\label{Eq:space_heom_vec}
|\Psi\rangle = \sum_{{\bf{n}},q,q'} \rho_{{\bf{n}}}(q,q') |q\rangle|\tilde{q}'\rangle |\bf{n}\rangle \;,
\end{equation}
where $|\bf{n}\rangle$ defines the state of the effective modes, with ${\bf{n}}=\{n_1,n_2,\cdots,n_{mk},\cdots \}$. 
The states $|q\rangle \in \mathcal{H}_S$ and $|\tilde{q}'\rangle \in \tilde{\mathcal{H}}_S$ belong to the Hilbert space of the system, and its corresponding fictitious twin space, respectively, as formulated by thermo-field theory.\cite{suzuki85}
This representation of a density matrix as a state vector is also known as purification in quantum computing.\cite{verstraete04,feiguin05,nielsen10}

The coefficients $\rho_{{\bf{n}}}(q,q') = \langle q |\hat{\rho}_{\bf{n}} |q' \rangle$, introduced by eq~\ref{Eq:space_heom_vec}, are the matrix elements of the density operator in the system Liouville space. 
The coefficient ${\rho}_{\vec{\bf 0}}(q,q')$, with $\vec{\bf 0} \equiv\{0,0,\cdots,0\}$, corresponds to the reduced density operator (RDO), defined as the partial trace of the total density operator $\rho_T(t)$ over the bath degrees of freedom: $\rho(t) = {\rm Tr}_B[\rho_T(t)]$.  This RDO describes the evolution of the reduced system.

With the twin-space formulation of thermo-field theory, the HEOM can be written as a time-dependent Schr\"odinger-like equation for the evolution of $|\Psi\rangle$,\cite{borrelli19,ke22}
\begin{equation}\label{Eq:twin_HEOM0}
\frac{d|\Psi\rangle}{dt} = -i \mathbb{H} |\Psi\rangle \,,
\end{equation}
where the effective Hamiltonian $\mathbb{H}$ is 
\begin{equation}\label{Eq:effH_all}
\begin{split}
\mathbb{H} = \hat{H}_S - \tilde{H}_S 
&- i \sum_{mk} v_{k} \hat{b}^\dag_{mk} \hat{b}_{mk}
+ \sum_{mk} \hat{A}_m \left(\sqrt{r_{k}}\hat{b}_{mk}+
\frac{d_{k}}{\sqrt{r_{k}}} \hat{b}^\dag_{mk}\right) \\
&-  \sum_{mk} \tilde{A}_m \left(\sqrt{r_{k}}\hat{b}_{mk}+\frac{{d}^*_{k}}{\sqrt{r_{k}} }\hat{b}^\dag_{mk}\right) \;\;.
\end{split}
\end{equation}
Operators with hats $(\hat{\mathcal{O}})$ act on the system Hilbert space, while tilded operators $(\tilde{\mathcal{O}})$ act on the fictitious space: $\hat{O} |q\rangle|\tilde{q}'\rangle \equiv O \otimes I |q\rangle|\tilde{q}'\rangle$, and $\tilde{O} |q\rangle|\tilde{q}'\rangle \equiv I \otimes O^T |q\rangle|\tilde{q}'\rangle$, where $I$ is the identity operator. 
Creation and annihilation operators for the effective modes satisfy
\begin{equation}
\begin{split}
\hat{b}^\dag_{mk} \, |n_1,\cdots,n_{mk},\cdots\rangle &= \sqrt{n_{mk}+1}\, |n_1,\cdots,n_{mk}+1,\cdots\rangle \,,\\
\hat{b}_{mk}\, |n_1,\cdots,n_{mk},\cdots\rangle &= \sqrt{n_{mk}}\, |n_1,\cdots,n_{mk}-1,\cdots\rangle \;\;.
\end{split}
\end{equation}
The scaling factor $r_{k}$ allows flexible definitions of the operators (if $[\hat{b}_{mk},\hat{b}^\dag_{mk}]=1$ then $[\sqrt{r_{k}}\hat{b}_{mk},\frac{\hat{b}^\dag_{mk}}{\sqrt{r_{k}}}]=1$ still satisfies the commutation relation). 
For this work, we use $r_{k} = |d_{k}|$, corresponding to the efficient filtering algorithm by Shi {\em et al.}\cite{shi09b,shi09c}

In practice, the number of effective modes $K$ ({\em i.e.}, the number of terms in the decomposition in eq~\ref{Eq:Ct_SOP}) and the Fock space size $L$ defining the truncation of the basis for each effective mode, with $n_{mk}\leq L$, need to be limited. We ensure convergence by increasing both parameters until stable results are achieved. For our model systems in this work, $K \leq 5$ is sufficient. However, for more complex spectral densities or lower temperatures, larger values of $K$ might be required, posing computational challenges. Those challenges could be addressed by recent advances, such as Tensor-Train methods,~\cite{shi18,yan21a,ke22,ke23} and more advanced reservoir correlation function decomposition schemes (e.g., Pad\'e spectrum decomposition,\cite{hu10,hu11} Fano spectrum decomposition (FSD),\cite{cui19,zhang20} barycentric spectrum decomposition (BSD),\cite{xu22,dan23} Prony fitting decomposition (PFD),\cite{chen22} and others.\cite{ye17,tian12,popescu15,nakamura18,rahman19,tang15,nakamura18,ikeda20,erpenbeck18}).

\subsection{The Projected Propagator}\label{sec:proj_propagator}
In this section, we present the general method for calculating the propagator $G(t)$, introduced by eq~\ref{Eq:qProp}, using the HEOM method.

As described in section~\ref{sec:secheom}, the HEOM method casts the effect of the bath on the system in terms of a small number of effective modes, thereby significantly reducing the computational cost. 
However, the Hilbert space of the state vector $|\Psi\rangle$ can still be very large. Therefore, we further reduce the dimensionality of the problem by projecting $|\Psi\rangle$ onto a smaller subspace, as described in Section \ref{Sec: qb_state_evolve}. The projected state vector is then encoded into the state vector of qubits,
\begin{equation}\label{Eq:encode}
 |\Phi\rangle = \mathcal{M}( \mathcal{P} |\Psi\rangle )  \;,
\end{equation}
where $\mathcal{P}$ is the projection operator that projects the state $|\Psi\rangle$ onto a subspace corresponding to the relevant physical quantities. $\mathcal{M}$ maps the projected subspace state $\mathcal{P} |\Psi\rangle$ to the qubit state vector $|\Phi\rangle$. 

\begin{figure}[!h]
\includegraphics[width=0.8\linewidth]{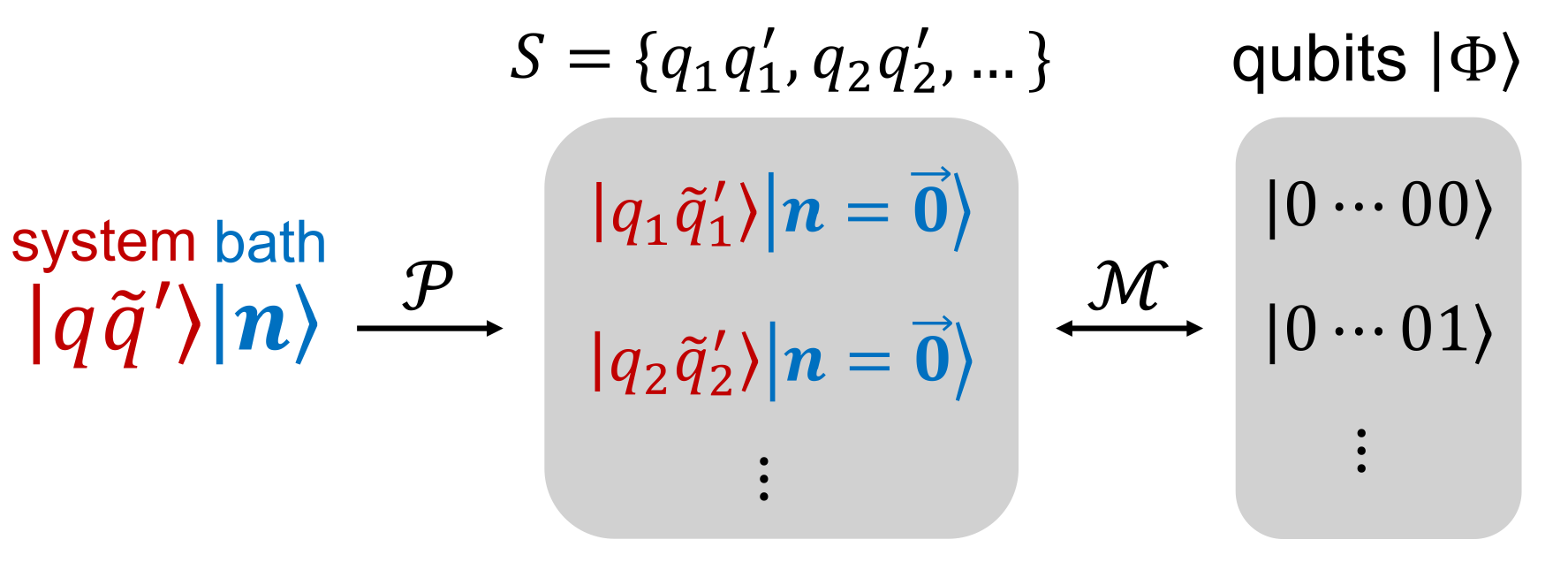}
\caption{Illustration of the projection and encoding processes. 
The HEOM computational space is projected by $\mathcal{P}$ onto a subspace where all effective modes are in state $|0\rangle$. The subspace is defined by the set $S$, which includes states relevant to the physical quantities of interest. 
The states within this subspace are further encoded into the qubit state vector $|\Phi\rangle$, with each qubit in either $|0\rangle$ or $|1\rangle$.
States in different spaces are marked with different colors in the figure: red represents the system Hilbert space, blue corresponds to the effective mode (bath) space, and black denotes the qubit state space.}
\label{Fig:proj_statevec}
\vspace{1em}
\end{figure}

We use the following projection operator:
\begin{equation}\label{Eq:ProjOpr}
   \mathcal{P} = \sum_{qq' \in S} |q \tilde{q}'\rangle |\vec{\bf 0}\rangle \langle q \tilde{q}'| \langle \vec{\bf 0} | \;, 
\end{equation}
where $S$ is the subspace containing all the states relevant to the physical quantities of interest.
$\vec{\bf 0} \equiv\{0,0,\cdots,0\}$ corresponds to all effective modes being in the 0 state, which is associated with the space of the reduced density operator.
The mapping $\mathcal{M}$ then encodes states in this subspace into qubit states, as follows:
\begin{equation}
    \mathcal{M}: |q_j \tilde{q}_j'\rangle |\vec{\bf 0}\rangle \to |j_n,...,j_1\rangle \,,
\end{equation}
where $q_j q_j'$ is the $j$-th element in $S$, and $(j_n,...,j_1)$ represents the $n$-bit binary form of the integer $j$. Here, $n=\log_2 N_S$ gives the number of qubits required to encode the $N_S$ relevant states in $S$. 
Consequently, the state $|q_j \tilde{q}_j'\rangle |\vec{\bf 0}\rangle$ is encoded as the qubit state $|j_n,...,j_1\rangle$. 
A schematic representation of the projection and encoding processes is shown in Figure~\ref{Fig:proj_statevec}.

To derive the expression for the propagator in eq~\ref{Eq:qProp}, we first write the formal solution of the HEOM. By integrating eq \ref{Eq:twin_HEOM0}, we obtain:
\begin{equation}
    |\Psi(t)\rangle = e^{-i \mathbb{H} t}  |\Psi(0)\rangle \;\;,
\end{equation}
then performing the Projection $\mathcal{P}$ and mapping $\mathcal{M}$, we have 
\begin{equation}\label{Eq:MPPsit}
    \mathcal{M}\mathcal{P}|\Psi(t)\rangle = \mathcal{M}\mathcal{P} e^{-i \mathbb{H} t} \mathcal{P} \mathcal{M}^\dag
 \mathcal{M}\mathcal{P}|\Psi(0)\rangle \;\;.
\end{equation}
Here, we utilize the properties $\mathcal{P}\mathcal{P} = \mathcal{P}$ and $\mathcal{M}^\dag \mathcal{M}$ being the identity operator in the projected subspace, which together yield $\mathcal{P} \mathcal{M}^\dag \mathcal{M}\mathcal{P} = \mathcal{P}$. We also impose the constraint $\mathcal{P}|\Psi(0)\rangle = |\Psi(0)\rangle$, a condition commonly used when selecting the Nakajima–Zwanzig–Mori projection operator in the derivation of the GQME.\cite{nakajima58,zwanzig60b,mori65}
By comparing the above expression with eq~\ref{Eq:qProp} and utilizing eq~\ref{Eq:encode},
the propagator in the reduced space has dimensions $N_S\times N_S$ and is given by:
\begin{equation}\label{Eq:Propagator}
G(t) = \mathcal{M} \mathcal{P} e^{-i \mathbb{H} t} \mathcal{P} \mathcal{M}^\dag \;,
\end{equation}
where $\mathbb{H}$ is the effective Hamiltonian of HEOM defined in eq~\ref{Eq:effH_all}. 
In this work, the propagator $G(t)$ is calculated on a classical computer. The process starts with solving the HEOM ({\em i.e.}, $e^{-i \mathbb{H} t}$ in eq~\ref{Eq:Propagator}), followed by applying the projection $\mathcal{P}$ onto the subspace corresponding to the quantity of interest. Finally, through the encoding process $\mathcal{M}$, the propagator is mapped onto its representation within the qubit state space.

It should be noted that the dynamics within the subspace $S$ are still numerically exact, in the sense that the numerical exact HEOM governs the evolution of the subspace elements. Simulations of physical quantities of interest require a suitable choice of projection operators to the corresponding subspace of those physical properties.

\subsection{The time-convolutionless 
Redfield equation within the rotating wave approximation}

The Lindblad-type quantum master equation is widely used in the field of quantum information science to model decoherence and open quantum system dynamics.~\cite{hu20,schlimgen21,schlimgen22b,schlimgen22,kamakari22,bacon01,cleve16,endo20,hu22,luo24} 
However, research indicates that the Lindblad equation, which rests on several rather restrictive assumptions, is only applicable within the semiclassical dynamics limit.\cite{koyanagi24,koyanagi24b,tanimura14,tanimura15} Therefore, its applicability should be carefully examined before using it to predict the dynamics of the open systems under consideration. 
In what follows, we perform such a test by comparing and contrasting the predictions of the Lindblad-type quantum master equation to those of the numerically exact HEOM for the model systems under consideration. 

For an open quantum system governed by the Hamiltonian given in eqs~\ref{Eq:Htot} and \ref{Eq:bath_coup_general}, the time-convolutionless (TCL) Redfield equation can be derived under the assumptions of
weak system-bath coupling and Markovian dynamics.
By further applying the rotating wave approximation (RWA), the equation takes the Lindblad form and is expressed as follows:
~\cite{breuer02,mohseni08,manzano20}
\begin{equation}\label{Eq:lindblad}
\frac{d\rho(t)}{dt} = -i [H_S+H_{LS},\rho(t)] 
+ \sum_\omega \sum_{m} \gamma_{m}(\omega) 
\left(A_m(\omega)\rho(t) A_m^\dag(\omega)
-\frac{1}{2}\{ A_m^\dag(\omega)A_m(\omega),\rho(t) \}
\right) .
\end{equation}
For simplicity, we will hereafter also refer to this equation as the Lindblad equation.
In this equation, $\rho(t)$ is the reduced density operator that describes the state of the system and $[\cdot, \cdot]$ and $\{\cdot,\cdot \}$ correspond to the commutator and anti-commutator, respectively. $H_{LS}$ is the Lamb shift Hamiltonian given by eq~\ref{eq:HLS},\cite{breuer02} which accounts for 
the energy shift induced by the interaction with the environment, and commutes with the system Hamiltonian ({\em i.e.}, $[H_S,H_{LS}]=0$).  
$\{ \gamma_{m}(\omega) \}$ and $\{ A_m(\omega) \}$ are damping rate coefficients and system jump operators, respectively, which are given by:
\begin{align}
\gamma_{m}(\omega) &= \int_{-\infty}^{\infty} dt \, e^{i\omega t} C_{m}(t) \;,\\
A_m(\omega) &= \sum_{\epsilon'-\epsilon = \omega} \langle \epsilon |A_m|\epsilon'\rangle |\epsilon \rangle\langle \epsilon'| \;,\\
H_{LS} &= \sum_\omega \sum_{m}S_{m}(\omega) A_m^\dag(\omega)A_m(\omega) \;,
\label{eq:HLS}
\end{align}
Here, $\omega = \epsilon'-\epsilon$, where $\epsilon$ and $\epsilon'$ are the eigenvalues of $H_S$, corresponding to the eigenstates $|\epsilon\rangle$ and $|\epsilon'\rangle$, respectively, and $C_{m}(t)$ is the bath correlation function defined in eq~\ref{Eq:Ct_jome}. Here we explicitly reintroduce the subscript $m$ to indicate that it corresponds to the correlation function of the $m$-th bath, which is coupled to the system through the operator $A_m$.
Finally, $S_{m}(\omega)$ is given by:
\begin{align}
S_{m}(\omega) &= \frac{1}{2i}(\Gamma_{m}(\omega)-\Gamma_{m}^*(\omega)) \;,\\
\Gamma_{m}(\omega) &= \int_0^{\infty} dt\, e^{i\omega t} C_m(t) \;.
\end{align}

Importantly, the validity of the Lindblad equation relies on three critical approximations.~\cite{breuer07,manzano20} First, the Born approximation assumes weak system-bath coupling within the framework of second-order perturbation theory. Second, the Markovian approximation assumes that the timescale of bath fluctuations is much faster than the timescale of the system's damping. Third, the RWA assumes that rapidly oscillating terms, compared to the time scale of the system's dynamics can be neglected. 

Eq~\ref{Eq:lindblad} can also be solved using the quantum algorithm presented in section \ref{Sec: qHEOM_algorithm} by replacing the HEOM propagator in eq~\ref{Eq:Propagator} with the propagator based on the Lindblad equation. The Lindblad equation propagator can be computed using the quantum algorithm outlined in Ref.~\citenum{schlimgen22b}. The accuracy of the quantum algorithm depends on the accuracy of the equation of motion that underlies the propagator. Since circuit construction is the same once the propagator is obtained, the cost difference between implementing the HEOM and Lindblad equation using the quantum algorithm primarily lies in the propagator calculation, which is performed on a classical computer.

The popularity of the Lindblad equation can be traced back to the fact that the description of open quantum system dynamics becomes analytically tractable under those approximations. However, the restrictive nature of those assumptions requires validation, particularly when applied to ultrafast processes of molecular systems such as electronic charge and energy transfer where those assumptions might break down (see below).

\section{Model Systems for Charge and Energy Transfer}\label{Sec: ModelSystem}

\begin{figure}[h]
\begin{center}
\includegraphics[width=1\linewidth]{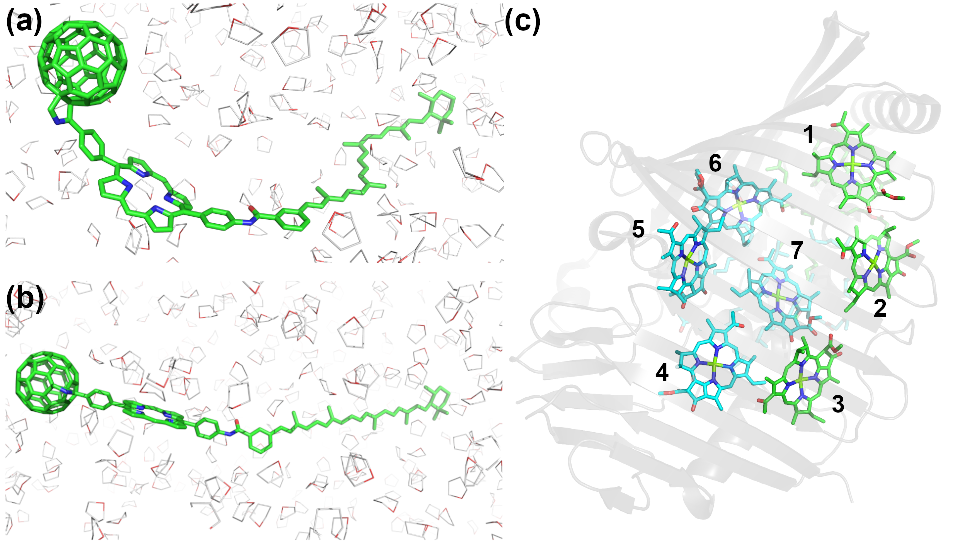}
\end{center}
\caption{Schematic representation of model systems for electron and energy transfer.
(a) Bent conformation and (b) linear conformation of the  carotenoid-porphyrin-\ce{C60}
(\ce{CPC60}) molecular triad dissolved in tetrahydrofuran. 
Charge transfer can occur after photoexcitation of the molecular triad.
(c) The FMO complex, 
where two pathways exist for excitation energy transfer: starts at site 1, transferring through site 2 to site 3, or starts at site 6 and passes through sites 5, 7, and 4, then reaching site 3.\cite{ishizaki09b}}
\label{Fig:model_all}
\end{figure}

In what follows, we will demonstrate the accuracy and utility of the quantum algorithm described in section~\ref{Sec: qHEOM_algorithm} 
by applying it to models of charge transfer in a solvated molecular triad and excitation energy transfer in the FMO complex. 
Figure~\ref{Fig:model_all} provides schematic representations of the model systems. For those model systems, 
the system corresponds to the electronic degrees of freedom, while the nuclear degrees of freedom play the role of the bath. 
Furthermore, the system-bath coupling plays a crucial role in determining the rates of charge and energy transfer in both cases. 
Below, we outline the  Hamiltonians for the two model systems under consideration.

\subsection{Model Hamiltonian for Charge Transfer in a Molecular Triad}

The first model we examine is photoinduced charge transfer within the carotenoid-porphyrin-\ce{C60} (\ce{CPC60}) molecular triad dissolved in tetrahydrofuran, which has been recently investigated extensively using a variety of semiclassical rate theories based on inputs from molecular dynamics simulations and time-dependent density functional theory (TDDFT) calculations.\cite{
sun18,tong20,han20,hu20b,brian21,brian21b,brian21c,hu22b}

Upon photoexcitation, the \ce{CPC60} transitions from its ground state to the porphyrin-localized excited $\pi\pi^*$ state, \ce{CP^*C60}. The system subsequently undergoes electron transfer from the porphyrin to the C$_{60}$, to form the so-called CT1 state, \ce{CP^+C^-_60}. 
Further hole transfer from the porphyrin to the carotene subsequently leads to the formation of the so-called CT2 state,  \ce{C^+PC^-_60} state.\cite{sun18} 
This sequence of events from photoexcitation to the formation of CT2 can then be summarized as follows: 
\begin{equation}
\ce{CPC60 ->T[$h\nu$] CP^*C60($\pi\pi^*$) -> CP^+C^-_60(${\rm CT1}$) -> C^+PC^-_60(${\rm CT2}$)}.
\end{equation}
Two dominant characteristic conformations were reported when \ce{CPC60} is dissolved in liquid tetrahydrofuran.\cite{sun18} 
These conformations are denoted \emph{bent} and \emph{linear} and are shown in Figure~\ref{Fig:model_all}(a) and (b), respectively. Importantly, the charge transfer rates are conformation-dependent.

In what follows, we will focus on the rate of the $\pi\pi^* \rightarrow {\rm CT1}$ charge transfer process. To this end, we map the system onto a spin-boson model with a Hamiltonian of the following form:\cite{tong20}
\begin{equation}\label{Eq:SpinBosonHamil}
{H}_T = V\sigma_x + E_0 \sigma_z + 
\sum_j \frac{p^2_j}{2 M_j}+\frac{1}{2} M_j \omega^2_j(x_j-\frac{c_j}{M_j \omega_j^2}\sigma_z)^2  \;.
\end{equation}
Here, the donor state $\pi\pi^*$ is represented as $|D\rangle=[1,0]^T$, and the acceptor state CT1 as $|A\rangle=[0,1]^T$. The electronic coupling between these states is denoted by $V$, while $E_0$ represents the energy difference between these two electronic states. Therefore, considering eqs~\ref{Eq:Htot} and \ref{Eq:bath_coup_general}, the subsystem Hamiltonian is ${H}_S = V\sigma_x + E_0 \sigma_z$, with no dependence on $m$, $A_m = \sigma_z$. The intramolecular and intermolecular (solvent) nuclear degrees of freedom are treated as a harmonic bath linearly coupled to the system.
The initial state is chosen to be
\begin{equation}\label{Eq:ini_SBM}
    \rho_T(t=0) = |D\rangle\langle D| \otimes \frac{e^{-\beta H_B}}{{\rm Tr}_B[e^{-\beta H_B}]} \;,
\end{equation}
where $\rho_T$ is the total density operator, and the bath DOFs are at thermal equilibrium.
The coupling strength $(\eta)$ and the width $\omega_c$ of the spectral density in eq~\ref{Eq:Debye} are chosen to match the reorganization energy and the width of the spectral density from Ref.~\citenum{tong20}. Table \ref{Tab:para} lists the parameters for the $\pi\pi^* \rightarrow {\rm CT1}$ charge transfer process corresponding to the bent and linear conformations of the molecular triad in tetrahydrofuran.

\begin{table}[h]
\begin{tabular}{m{4cm}<{\centering}|m{5cm}<{\centering}|m{5cm}<{\centering}}
\toprule [2pt]
Parameter &     Bent & Linear   \\ 
\midrule [1pt]
$V$ & $2.4\times 10^{-2}$~eV &  $9.0 \times 10^{-3}$~eV   \\
$E_0$ & $0.507$~eV &  $0.236$~eV  \\ 
$\eta$ & $0.2565$~eV & $0.318$~eV  \\
$\omega_c$ & 25~${\rm cm}^{-1}$ & 25~${\rm cm}^{-1}$  \\
T & 300~K & 300~K  \\
\bottomrule [2pt]
\end{tabular}
\caption{The model parameters of the spin-boson model correspond to the 
$\pi\pi^* \rightarrow {\rm CT1}$ charge transfer process for the bent and linear conformations of the \ce{CPC60} in tetrahydrofuran solution.\cite{tong20}}
\label{Tab:para}
\vspace{7em}
\end{table}

The charge transfer dynamics in \ce{CPC60} were simulated via the numerically exact HEOM framework, eq~\ref{Eq:twin_HEOM0}, as well as via the approximate Lindblad equation, eq~\ref{Eq:lindblad}. In the HEOM simulations, the initial state was set according to eq~\ref{Eq:ini_SBM}, with $|\Psi_{MT}(t=0)\rangle = |D\rangle |\tilde{D}\rangle|\bf 0\rangle$. The effective Hamiltonian for HEOM is
\begin{equation}\label{Eq:effH_SBM}
\mathbb{H}_{MT} = \hat{H}_S - \tilde{H}_S 
- i \sum_{k=1}^K v_k \hat{b}^\dag_k \hat{b}_k
+ \hat{\sigma}_z\sum_{k=1}^K  \left(\sqrt{r_k}\hat{b}_k+
\frac{d_k}{\sqrt{r_k}} \hat{b}^\dag_k\right)
- \tilde{\sigma}_z \sum_{k=1}^K \left(\sqrt{r_k}\hat{b}_k+\frac{{d}^*_k}{\sqrt{r_k} }\hat{b}^\dag_k\right)\;,
\end{equation}
where ``MT" denotes the molecular triad model. 

For the projection operator in the form of eq~\ref{Eq:ProjOpr}, we consider the full reduced density matrix as the quantity of interest, with the subspace $S=\{DD,DA,AD,AA\}$. This four-state subspace can be encoded in two qubits, resulting in a 3-qubit circuit with the SVD dilation method. For charge transfer in a molecular triad, where only electronic state populations are of interest, the subspace $S$ can be simplified to $S=\{DD,AA\}$, yielding a 2-qubit circuit. Later, we will show that this population-only subspace provides more accurate NISQ simulation results. 

Note that the dynamics within subspace $S$ remain numerically exact, and the limitation imposed by the subspace is that the quantum computer can only simulate the physical quantities within $S$. For $S=\{DD,DA,AD,AA\}$, the quantum circuit captures the full reduced density matrix dynamics. However, with the population-only subspace $S=\{DD,AA\}$, the circuit cannot simulate the coherences ($\langle A|\hat{\rho}|D\rangle$, $\langle D|\hat{\rho}|A\rangle$). Such subspace selection has also been adopted in recent studies.\cite{wang23,lyu24}

\subsection{Model Hamiltonian for Energy Transfer in the FMO Complex}\label{Sec: FMO_Model}

The Fenna-Matthews-Olson (FMO) complex is a well-characterized light-harvesting system\cite{fenna75,li97,ishizaki09a,ishizaki09b,moix11,sarovar11,schulze16} 
that serves as a quantum conduit, directing excitation energy from the light-harvesting antenna to the reaction center.\cite{engel07,scholes17} This process involves exciton transfer between the seven bacteriochlorophyll (Bchl) chromophores comprising the FMO complex.
Figure~\ref{Fig:model_all}(c) provides a schematic representation of the system. 

FMO is often described in terms of a Frenkel exciton Hamiltonian:
\begin{equation}\label{Eq:FMO_H}
\begin{split}
    H_{FMO} =& \sum_{m=1}^N \epsilon_m |m\rangle\langle m| + 
    \sum_{m<n} J_{mn} (|m\rangle\langle n|+|n\rangle\langle m|) \\ 
    &+ \sum_{m=1}^N \sum_{j=1}^{N^m_{b}} \left( \frac{p_{mj}^2}{2M_{mj}} + \frac{1}{2} M_{mj} \omega_{mj}^2 x_{mj}^2 - c_{mj} x_{mj}  |m\rangle\langle m|  \right) \;.
\end{split}
\end{equation}
Here, $|m\rangle$ represents the excited state of site $m$, which corresponds to locally exciting the $m$-th BChl chromophore. 
$\epsilon_m$ is the site excitation energy, and $J_{mn}$ denotes the dipolar coupling between sites $m$ and $n$. 
Each site is coupled to its phonon bath, with $N^m_{b}$ phonon modes per bath. The parameter $c_{mj}$ defines the coupling strength of phonon mode $j$ to site $m$. 
Following eqs.~\ref{Eq:Htot} and \ref{Eq:bath_coup_general}, we identify $H_S = \sum_{m=1}^N \epsilon_m |m\rangle\langle m| + \sum_{m<n} J_{mn} (|m\rangle\langle n|+|n\rangle\langle m|)$ as the system Hamiltonian, and $A_m = |m\rangle\langle m|$. 

In this study, we use the seven-site model Hamiltonian ($N=7$) for the FMO complex, with $\epsilon_m$ and $J_{mn}$ values obtained from Moix. \emph{et al}.\cite{moix11} 
The matrix representation of $H_S$ (in units of ${\rm cm}^{-1}$) is
\begin{equation}\label{Eq:Hs_FMO_mat}
    H_S = \left( 
    \begin{array}{ccccccc}
       310.0  & -97.9 &   5.5 &   -5.8  &  6.7  &  -12.1  &  -10.3 \\
  -97.9  &   230.0  & 30.1  &  7.3  &  2.0   &  11.5  &   4.8 \\
   5.5   &   30.1  &  0   &   -58.8 & -1.5  &  -9.6   &   4.7 \\
  -5.8   &   7.3  &  -58.8 &   180.0 & -64.9 &  -17.4 &   -64.4 \\
   6.7   &   2.0  &  -1.5  &  -64.9 &  405.0  & 89.0   & -6.4 \\
  -12.1   &  11.5 &  -9.6  &  -17.4 &  89.0  &  320.0 &   31.7 \\
  -10.3  &   4.8  &   4.7 &   -64.4 & -6.4   &  31.7   &  270.0 
    \end{array} \right) \;\;.
\end{equation}
For the bath, each site couples to an identical bath ({\em i.e.}, $c_{mj}\equiv c_j$) and follows a Debye spectral density with parameters $\eta = 70~{\rm cm}^{-1}$ and $\omega_c^{-1} = 50$~fs.\cite{ishizaki09b,zhu11,shi18}

Excitation energy transfer within the FMO complex follows two primary pathways: either starting at site 1 and passing through site 2 to site 3, or starting at site 6 and passing through sites 5, 7, and 4, to reach site 3.\cite{ishizaki09b} 
We focus on the first pathway, where the initial state is $|m=1\rangle$ and the phonon bath is at thermal equilibrium: 
\begin{equation}\label{Eq:ini_FMO}
    \rho_T(t=0) = |1\rangle\langle 1| \otimes \frac{e^{-\beta H_B}}{{\rm Tr}_B[e^{-\beta H_B}]} \;\;.
\end{equation}
In this setup, the initial HEOM state vector is $|\Psi_{FMO}(t=0)\rangle = |1\rangle |\tilde{1}\rangle|\bf 0\rangle$, and effective Hamiltonian is expressed as
\begin{equation}\label{Eq:effH_FMO}
\begin{split}
\mathbb{H}_{FMO} = \hat{H}_S - \tilde{H}_S 
&- i \sum_{m=1}^N\sum_{k=1}^K v_k \hat{b}^\dag_{mk} \hat{b}_{mk}
+ \sum_{m=1}^N\sum_{k=1}^K |m\rangle\langle m| \left(\sqrt{r_k}\hat{b}_{mk}+
\frac{d_k}{\sqrt{r_k}} \hat{b}^\dag_{mk}\right) \\
&-  \sum_{m=1}^N \sum_{k=1}^K |\tilde{m}\rangle\langle \tilde{m}|  \left(\sqrt{r_k}\hat{b}_{mk}+\frac{{d}^*_k}{\sqrt{r_k} }\hat{b}^\dag_{mk}\right) \;\;,
\end{split}
\end{equation}
where ``FMO" stands for the FMO complex. 

For the projection operator in the FMO complex model, we measure populations at sites 1, 2, 3, and 6, capturing the primary excitation pathway 
$1 \to 2 \to 3$.\cite{ishizaki09,moix11,hu22} Therefore, the projection operator is chosen as in eq~\ref{Eq:ProjOpr} with the projected subspace $S = \{11, 22, 33, 66\}$, which, after dilation, results in a 3-qubit circuit. This subspace can be decomposed into three smaller subspaces: $S_1=\{11,22\}$, $S_2=\{11,33\}$, and $S_3=\{11,66\}$. Notably, the constraint $\mathcal{P}|\Psi(0)\rangle = |\Psi(0)\rangle$ imposed on the projection operator (i.e. as shown below eq~\ref{Eq:MPPsit}) applies to these subspaces.

\section{Results}\label{Sec: result}

\begin{figure}[!h]
\begin{center}
\includegraphics[width=1\linewidth]{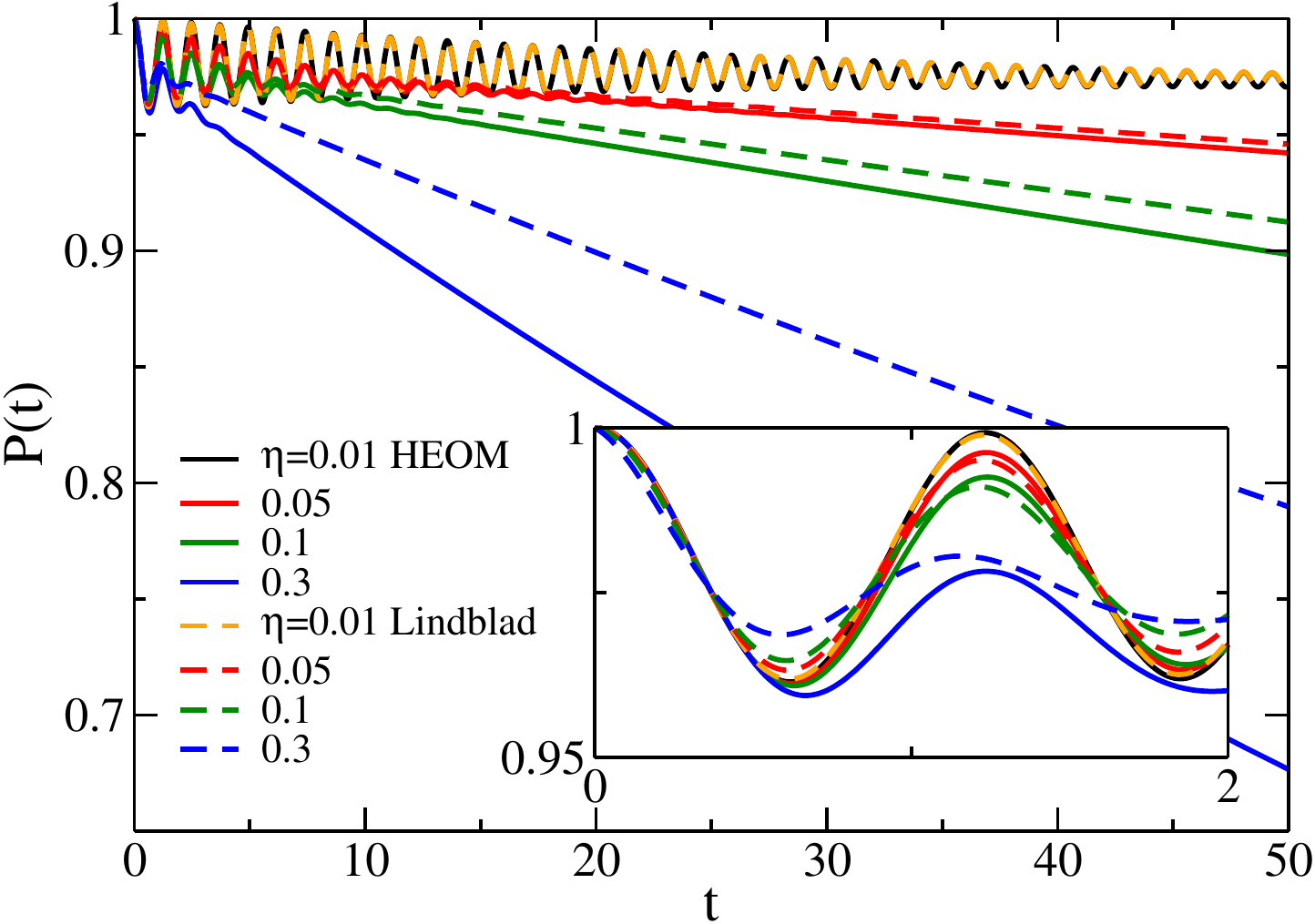}
\end{center}
\caption{Population dynamics $P(t)$ of the donor state in the spin-boson model across various system-bath coupling strengths $\eta$, computed using numerically exact HEOM (solid lines) and the Lindblad equation (dashed lines). 
Simulation parameters: $V=0.5, E_0=2.5, \beta=1, \omega_c=1$.
The inset highlights the transient dynamics in detail.}
\label{Fig:heom_vs_lindblad0}
\end{figure}

To assess the applicability of the Lindblad equation, we first tested a general spin-boson model (eq~\ref{Eq:SpinBosonHamil}) with tunable parameters. Figure~\ref{Fig:heom_vs_lindblad0} compares the dynamics of the donor state population at different system-bath coupling strengths $\eta$, as obtained via the Lindblad quantum master equation vs. the numerically exact HEOM method (the values of the remaining model parameters are set at $V=0.5$, $E_0=2.5$, $\beta=1$, and $\omega_c=1$). The inset provides a detailed view of the short-time dynamics. As expected, the Lindblad quantum master equation is seen to be accurate when $\eta$ is sufficiently small (the weak system-bath coupling limit). Increasing the value of $\eta$ beyond the weak coupling limit ($\eta \sim 0.01-0.05$), the Lindblad quantum master equation is found to overestimate the population relaxation rate at short times and underestimate it at longer times.  

It should also be noted that the dynamical range of validity of the Lindblad quantum master equation corresponds to a weak damping regime, and therefore gives rise to donor population dynamics that are pronouncedly coherent (oscillatory) and dominated by the electronic coupling $V$. 
While damping increasingly dominates the dynamics with increasing $\eta$, the regime where the dynamics are truly incoherent lies outside the range of validity of the Lindblad quantum master equation.
The breakdown of the Lindblad equation at strong system-environment coupling $\eta$ is consistent with the fact that it relies on the Born-Markov approximation.\cite{breuer02}
As noted in previous studies, perturbative methods become invalid when the perturbation strength is strong.\cite{jiang02,mavros14,xu18a,dan22}
The results in Figure~\ref{Fig:heom_vs_lindblad0} clearly demonstrate that the Lindblad quantum master equation is only applicable within the coherent dynamics regime ($\eta\ll V$). 
Detailed comparisons with the HEOM and Lindblad-typed TCL-Redfield equation can be found in Refs.\citenum{koyanagi24, koyanagi24b, tanimura15}.

\subsection{Charge Transfer in a Molecular Triad: Lindblad vs. HEOM}

\begin{figure}[h]
\begin{center}
\includegraphics[width=1\linewidth]{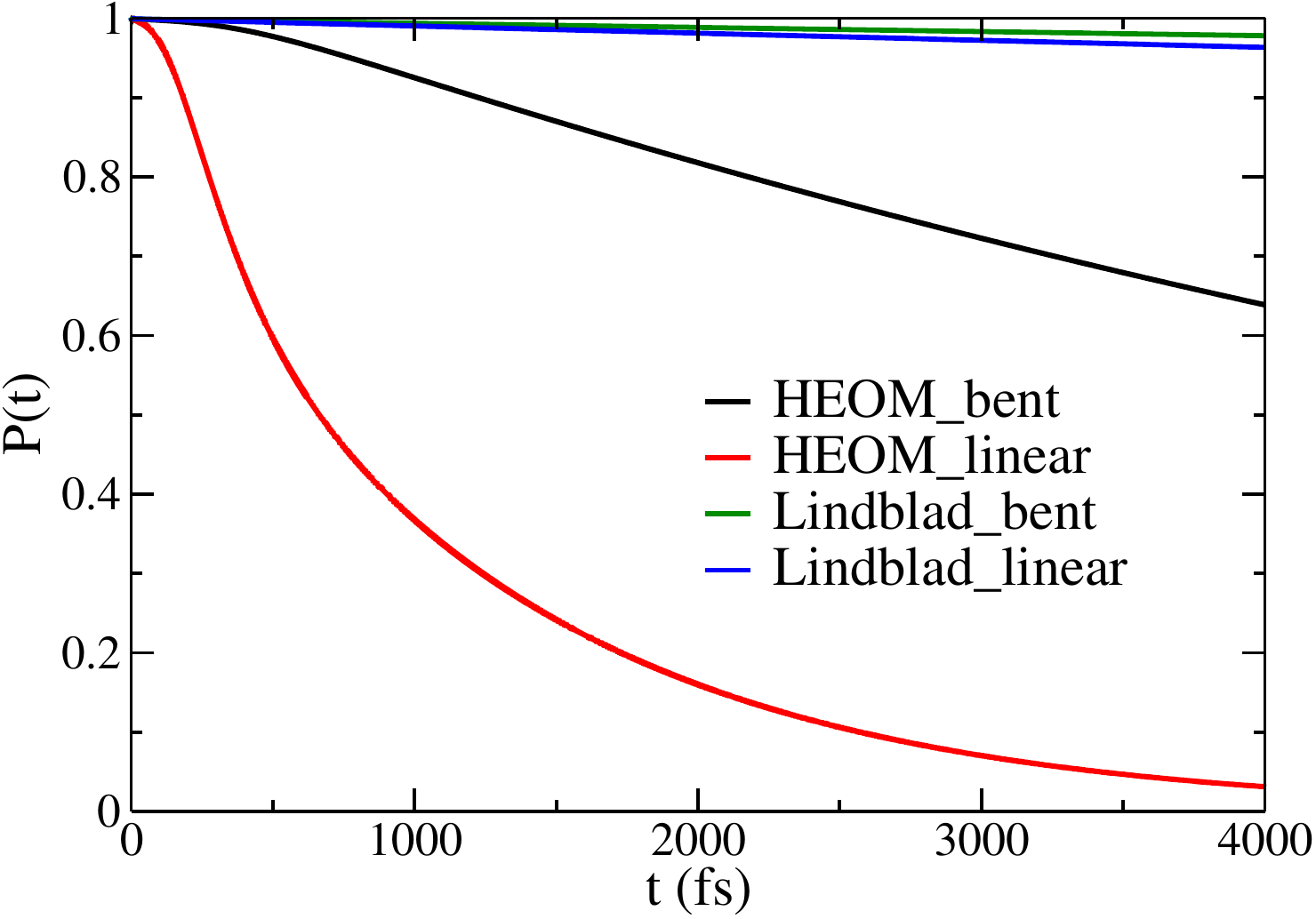}
\end{center}
\caption{$\pi\pi^*$ state population dynamics in the $\pi\pi^* \rightarrow {\rm CT1}$ photoinduced charge transfer process of \ce{CPC60} molecular triad dissolved in tetrahydrofuran. 
Here, the simulation results using the Lindblad equation and the numerically exact HEOM are shown for the two configurations of the \ce{CPC60} molecular triad (bent and linear). 
The parameters used in the simulations are in Table~\ref{Tab:para}.
}
\label{Fig:heom_vs_lindblad_CPC60}
\end{figure}

We now apply the Lindblad equation and HEOM to study electron transfer in the molecular triad. Figure~\ref{Fig:heom_vs_lindblad_CPC60} shows the donor state ($\pi\pi^*$) population dynamics for the $\pi\pi^* \rightarrow {\rm CT1}$ charge transfer process of \ce{CPC60} in tetrahydrofuran solution. Simulation parameters for the two \ce{CPC60} conformations (bent and linear) are provided in Table~\ref{Tab:para}.
\begin{table}[!h]
\begin{tabular}{m{3cm}<{\centering}|m{3cm}<{\centering}|m{3cm}<{\centering}|m{3cm}<{\centering}}
\toprule [2pt]
Rate constant &
HEOM  &  
Lindblad  & 
Marcus \\ 
\midrule [1pt]
 Bent &
 $1.24\times 10^{11}$~$\rm s^{-1}$ &
 $5.32\times 10^{9}$~$\rm s^{-1}$ &
 $1.19\times 10^{11}$~$\rm s^{-1}$ \\
Linear & $8.17\times 10^{11}$~$\rm s^{-1}$ &  $9.20\times 10^{9}$~$\rm s^{-1}$ &
 $1.13\times 10^{12}$~$\rm s^{-1}$ \\ 
\bottomrule [2pt]
\end{tabular}
\caption{Charge transfer rate constants calculated by the HEOM, the Lindblad equation, and the Marcus theory. 
The HEOM and Lindblad rate constants are obtained by exponential fitting of their respective $P(t)$ data in Figure~\ref{Fig:heom_vs_lindblad_CPC60}, in the range of $t=3000$ to $4000$~fs.}
\label{Tab:rate}
\end{table}

Inspection of the exact HEOM results shows that, for both bent and linear conformations, the population dynamics follow incoherent rate kinetic ($\eta \gg V$), with the long-time population of the $\pi\pi^*$ state exhibiting an
exponential decay. Notably, 
the $\pi\pi^* \rightarrow {\rm CT1}$ electron transfer is significantly faster in the linear conformation, which is consistent with previous studies.\cite{sun18} 
In contrast, the Lindblad quantum master equation predicts much slower charge transfer rates for both conformations, which is consistent with the fact that $\eta \gg V$ for this model (see Table \ref{Tab:para}). 
Despite these discrepancies, the Lindblad quantum master equation does capture some trends, such as the exponential decay in population dynamics and the faster charge transfer rate in the linear conformation.

To quantitatively illustrate the discrepancy between the predictions of the Lindblad quantum master equation and the exact results, we present the charge transfer rate constants predicted by HEOM and the Lindblad equation in Table~\ref{Tab:rate}. The rate constants are obtained from an exponential fit of the $P(t)$ data in Figure~\ref{Fig:heom_vs_lindblad_CPC60} over the time range $t = 3000$ to $4000$~fs. For comparison, we also include rate constants calculated using Marcus theory.\cite{marcus56,marcus85,marcus93} According to Marcus theory, the charge transfer rate constant $k_{Marcus}$ is given by:\cite{marcus56,marcus85,marcus93,tong20,nitzan06}
\begin{equation}
    k_{Marcus} = \frac{V^2}{\hbar} \sqrt{\frac{\pi}{\lambda k T}} \, e^{-\frac{(E_{DA}-\lambda)^2}{4\lambda kT}} \,,
\end{equation}
where $E_{DA} = 2E_0$ is the energy difference between the donor and acceptor states, and $\lambda$ is the reorganization energy. For a spin-boson model with Debye spectral density (eq~\ref{Eq:Debye}), $\lambda = 2\eta$.\cite{shi09c} The Marcus rate constants in Table~\ref{Tab:rate} are consistent with those reported by Tong \textit{et al}.\cite{tong20} Our $\pi \pi^*$ state dynamics results for the bent conformation also align with the cavity-free case in Ref.~\citenum{lyu24}, where minimal population change was observed within 5000~a.u. (120.9~fs); here, significant dynamics emerge only beyond the 1000~fs timescale.

As shown in Table~\ref{Tab:rate}, the exact HEOM rate constants closely match the Marcus rate constants, underscoring Marcus theory’s robustness in describing charge transfer in solution and validating the spin-boson model parameters in Table~\ref{Tab:para} and the HEOM results. In contrast, the Lindblad equation predicts significantly slower transfer rates, one to two orders of magnitude lower than the exact (HEOM) rates. This discrepancy is more pronounced for the linear conformation, with $k_{HEOM}/k_{Lindblad}$ values of 23.3 for the bent conformation and 88.8 for the linear conformation.

This outcome is expected. From Figure~\ref{Fig:heom_vs_lindblad0}, we observed that the Lindblad equation is only accurate when $\eta \ll V$, suggesting $\eta / V$ as a key parameter for assessing its validity. Table~\ref{Tab:para} shows $\eta / V = 10.7$ for the bent conformation and $\eta / V = 35.3$ for the linear conformation. These values indicate that the deviation between the Lindblad and exact rates ($k_{HEOM}/k_{Lindblad}$) is indeed proportional to $\eta / V$.

\subsection{Energy Transfer in the FMO Complex: Lindblad vs. HEOM}

\begin{figure}[!h]
\begin{center}
\includegraphics[width=1\linewidth]{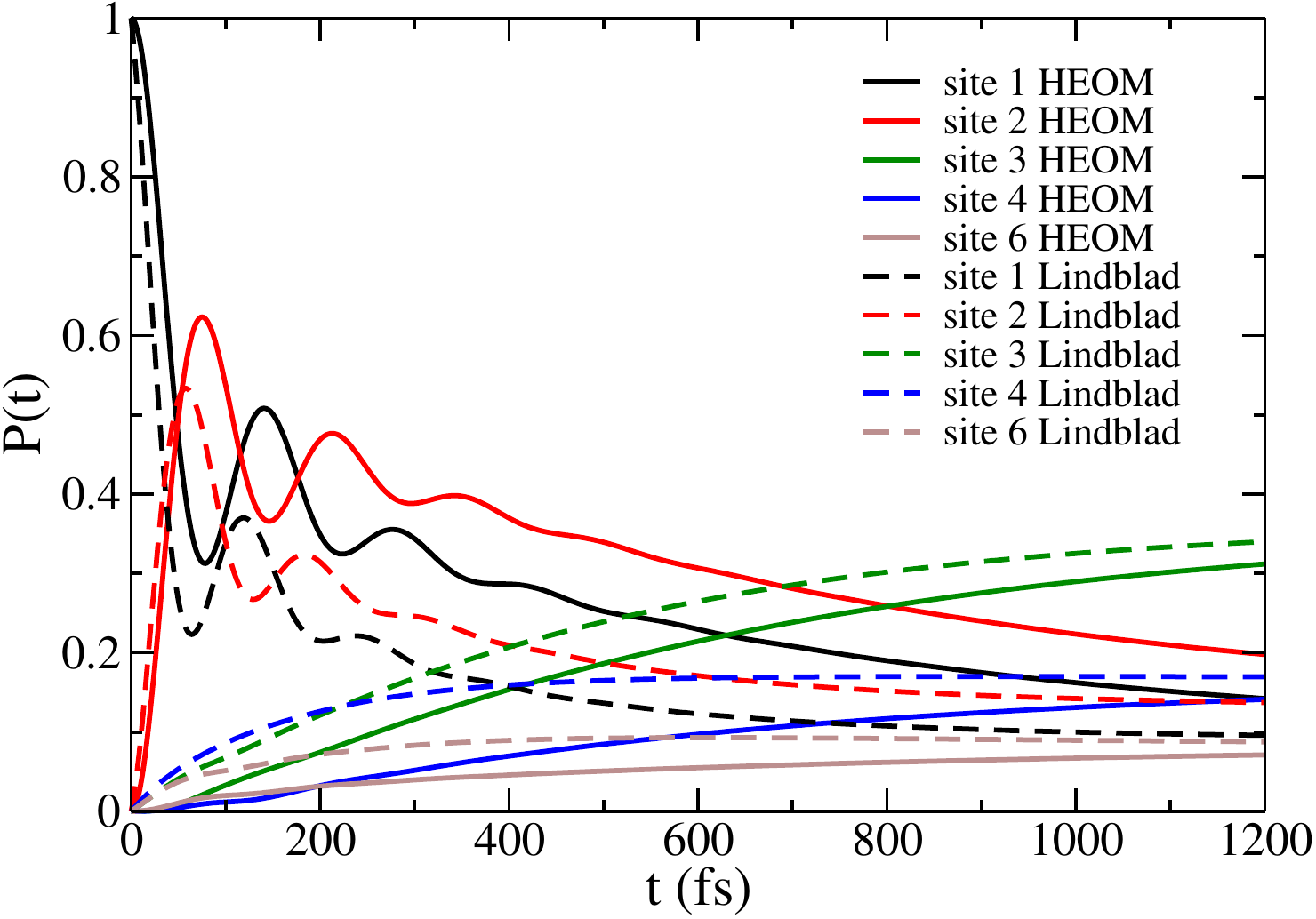}
\end{center}
\caption{Population dynamics of different sites in the FMO complex at 300~K with Debye spectral density. All parameters are defined in section~\ref{Sec: FMO_Model}, and dynamics are obtained using numerically exact HEOM (solid lines) and the Lindblad equation (dashed lines). The population is initialized at site 1.}
\label{Fig:heom_vs_lindblad_FMO}
\end{figure}

In Figure~\ref{Fig:heom_vs_lindblad_FMO}, we compare Lindblad dynamics with the numerically exact HEOM dynamics for excitation energy transfer in the FMO complex at 300~K. We present the population dynamics across different sites, with the initial population localized at site 1. Simulation parameters are specified in section~\ref{Sec: FMO_Model}.

The HEOM results in Figure~\ref{Fig:heom_vs_lindblad_FMO} show that within the transient dynamics regime ($t < 300$~fs), coherent oscillations occur between sites 1 and 2. 
These oscillations fade over longer timescales, transitioning into a rate kinetics regime at longer times.
As the population at site 3 grows, signifying energy transfer toward the terminal site and subsequently to the reaction center.\cite{moix11} This behavior stems from the fact that 
$\eta \sim J_{mn}$ in this case. 

The fact that $\eta \sim J_{mn}$ in FMO is also consistent with the observation that 
the Lindblad dynamics in Figure~\ref{Fig:heom_vs_lindblad_FMO} are in better agreement with the exact HEOM results (in comparison to the molecular triad case where $\eta \gg V$). 
However, the deviations between the Lindblad and HEOM results are rather large, with the former  
exhibiting faster short-time and slower long-time dynamics, as evidenced by the shallower slope of $P(t)$ for Lindblad results after $t=600$~fs. This outcome is consistent with the general trends shown in Figure~\ref{Fig:heom_vs_lindblad0}.

\subsection{Quantum Circuit Simulation Results}

\begin{figure}[!h]
\begin{center}
\includegraphics[width=0.55\linewidth]{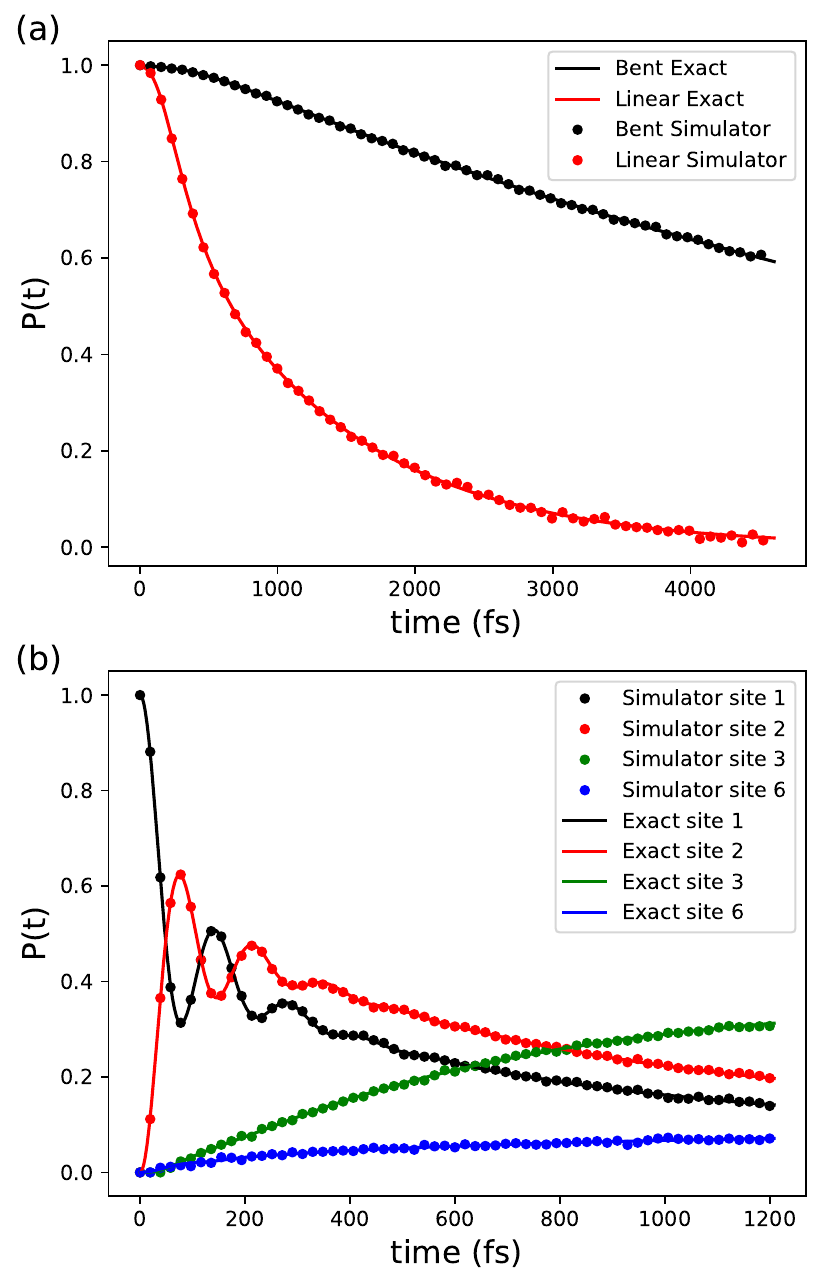}
\end{center}
\caption{Quantum circuit simulation of (a) $\pi\pi^*$ state population dynamics of the \ce{CPC60} molecular triad in tetrahydrofuran and (b) population dynamics of different sites in the FMO complex. The solid lines, labeled ``Exact", correspond to HEOM results from Figures~\ref{Fig:heom_vs_lindblad_CPC60} and \ref{Fig:heom_vs_lindblad_FMO}. Dotted points show quantum circuit results obtained via IBM QasmSimulator (Qiskit Aer\cite{qiskit24}), with $20000$ shots per time point.}
\label{fig:Aersimu}
\end{figure}

In this section, we report results obtained by applying the qHEOM algorithm described in Section \ref{Sec: qHEOM_algorithm} to simulate the charge and energy transfer dynamics in the two model systems under consideration on quantum circuits. Figure~\ref{fig:Aersimu} compares results obtained via qHEOM to results obtained via HEOM on a classical computer. 
The qHEOM results were obtained by utilizing IBM's noisy quantum circuit simulator, QasmSimulator from the Qiskit Aer package,~\cite{qiskit24} sampling $20000$ shots per time point. Figure~\ref{fig:Aersimu}(a) shows the dynamics of the $\pi \pi^*$ population in 
\ce{CPC60} based on the projection operator in eq~\ref{Eq:ProjOpr} which treats the full reduced electronic density matrix as the quantity of interest ({\em i.e.} projects onto the subspace $S=\{DD,DA,AD,AA\}$) and yields a 3-qubit circuit (with dilation). 
Figure~\ref{fig:Aersimu}(b) shows the energy transfer dynamics in the FMO complex, where we measure populations at sites 1, 2, 3, and 6 ({\em i.e.} the subspace $S = \{11, 22, 33, 66\}$). 

The excellent agreement between the qHEOM and HEOM results 
validates the accuracy of our quantum algorithm, as well as its ability to simulate non-unitary dynamics of open quantum systems on a unitary circuit. It also demonstrates the ability of qHEOM to accurately simulate non-Markovian dynamics beyond the range of applicability of the Lindblad quantum master equation. 

In this work, we focus on electronic energy and charge transfer dynamics in models of the molecular triad and the FMO complex. 
In those cases, the populations of electronic states constitute the quantity of interest. Importantly, the fact that the populations of electronic states constitute the quantity of interest does not imply that the coherences are left out or that the coherences do not impact the population dynamics. Furthermore, it should be emphasized that our quantum algorithm is not restricted to simulating population dynamics. 
For instance, in the case of the molecular triad model, the subspace $S=\{DD,DA,AD,AA\}$ includes all elements of the full reduced density matrix, not just populations. Similarly, for the FMO complex model, coherence dynamics can be incorporated by adjusting the projection subspace.

It should be noted that the construction of quantum circuits is based on the propagator $G(t)$, which, according to eq~\ref{Eq:Propagator}, is precomputed by solving the HEOM on a classical computer. The quantum circuit simulates the non-unitary process described by eq~\ref{Eq:qProp}, which can also be directly performed on a classical computer. Therefore, a direct comparison of computation times between classical and quantum implementations is not appropriate. Nonetheless, the purpose of this work is not to demonstrate quantum advantage but to illustrate how numerically exact HEOM can be implemented using quantum circuits based on unitary gates.

\begin{figure}[!tbh]
\begin{center}
\includegraphics[width=1\linewidth]{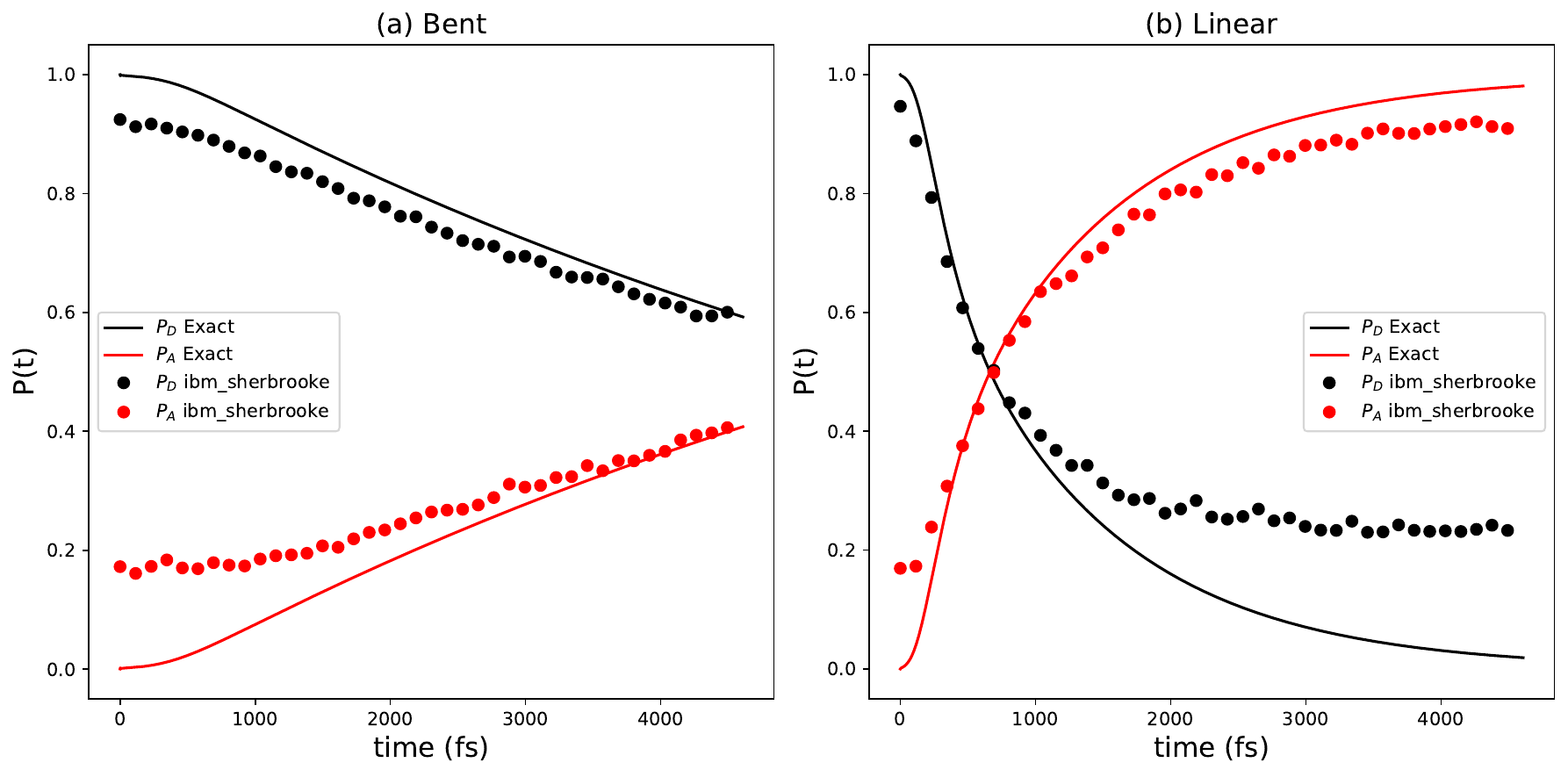}
\end{center}
\caption{Population dynamics of the \ce{CPC60} molecular triad: (a) Bent conformation, (b) Linear conformation. $P_D$ represents the population in the $\pi\pi^*$ (donor) state, and $P_A$ represents the population in the ${\rm CT1}$ (acceptor) state, with the projection subspace $S = \{DD, DA, AD, AA\}$. Solid lines show the exact HEOM results, while scatter points indicate quantum circuit results obtained from the IBM Sherbrooke quantum computer. Each time point was sampled with $20000$ shots. Error mitigation techniques, including dynamic decoupling (XX sequence) and 2-qubit Clifford gate twirling, were applied using Qiskit.
}
\label{fig:ibm3qb_CPC60}
\end{figure}

Next, we report the results obtained via qHEOM on NISQ devices. Figure~\ref{fig:ibm3qb_CPC60} shows the charge transfer dynamics in \ce{CPC60} as obtained by running qHEOM on the IBM Sherbrooke quantum computer, with the same projected subspace $S=\{DD,DA,AD,AA\}$ as in Figure~\ref{fig:Aersimu}(a). Populations for the donor $\pi\pi^*$ state [$P_D (t)$] and the acceptor ${\rm CT1}$ state [$P_A (t)$] were measured at 40 time points, each corresponding to an individual circuit sampled $20000$ times. Dynamic decoupling (XX sequence) and 2-qubit Clifford gate twirling error mitigation techniques were used within Qiskit. Circuit complexity remains consistent across time points, with an example circuit depth of 60 and 11 2-qubit gates shown in Supporting Information (SI).

Inspection of Figure~\ref{fig:ibm3qb_CPC60} reveals that while simulation on the NISQ device can reproduce the general dynamical behavior, noise gives rise to discrepancies compared with the exact results. These discrepancies exhibit notable patterns. First, larger errors occur when $P_D (t)$ or $P_A (t)$ approaches 0 (as the other population nears 1), particularly in the short-time region for the bent conformation and across short- and long-time regions for the linear conformation. 
This behavior is consistent with observations from previous studies,\cite{wang23,seneviratne24} which reported larger deviations in NISQ results at early times when the exact population of certain states is close to zero. Second, deviations are more pronounced when the exact population is near 0, compared to when it is near 1. In particular, $P_A (t)$ exhibits larger errors than $P_D (t)$ at short times for the bent conformation, and $P_D (t)$ shows greater errors than $P_A (t)$ at longer times for the linear conformation. This effect is attributed to noise in the sampling measurements. 

To explain this trend, we note that after running the circuit in Figure~\ref{Fig:qc_dilation}, the population for $|i\rangle$ with $ii\in S$ is retrieved by counting instances of $|i\tilde{i}{\bf 0}\rangle$ with the ancilla qubit in $|0\rangle$. If $N_i$ counts occur from $N_c$ total measurements, the population of state $|i\rangle$ is calculated as 
\begin{equation}\label{Eq:pop_qcount}
    P_i = \sigma_0 \sqrt{N_i/N_c} \;\;,
\end{equation}
where $\sigma_0$ is the largest singular value of the propagator (see eq~\ref{Eq:singval_Gt}).
Allowing for noise in $N_i$, $N_i=N_{exact}\pm N_{err}$, the error in $P_i$ can be estimated as follows: 
\begin{equation}\label{Eq:measure_error}
    P_i = \sigma_0 \sqrt{N_i/N_c} = P_{exact}(\sqrt{1 \pm N_{err}/N_{exact}}) \;\;.
\end{equation}
Thus, as $P_{exact}$ approaches 0, $N_{exact}$ is small, increasing the deviation of $(\sqrt{1 \pm N_{err}/N_{exact}})$ from 1.

Importantly, error mitigation does reduce deviations significantly, with results in the SI showing that the unmitigated NISQ data exhibits the same trends but with larger errors. The results in Figure~\ref{fig:ibm3qb_CPC60} surpass in accuracy previously reported results obtained on 3-qubit circuits in Ref.~\citenum{wang23}, where the Sz.-Nagy dilation method was employed to handle the non-unitary propagator $G(t)$. Here, SVD dilation with an efficient Walsh operator representation of $U_\Sigma$ reduced circuit depth, as shown in Table \ref{Tab:qc_depth_vs_Sz}. Thus, combining SVD dilation with Walsh operator implementation for $U_\Sigma$ reduces circuit complexity by more than a factor of 2 compared to the Sz.-Nagy approach.

\begin{table}
\caption{Circuit complexity for different dilation methods. ``SVD" refers to the SVD dilation approach shown in Figure~\ref{Fig:qc_dilation}, where the diagonal unitary operator $U_\Sigma$ is directly compiled, while ``SVD+Walsh" indicates the SVD dilation approach combined with the Walsh operator representation for $U_\Sigma$. The dilation is based on the propagator $G(t)$ at $t=2073.5$~fs for the linear conformation shown in Figure~\ref{fig:ibm3qb_CPC60}. Circuits are compiled to the basis gate set of the IBM Sherbrooke quantum computer ($X$, $\sqrt{X}$, $R_z$, and ECR) and adapted to its specific topology. The 2-qubit gate count reflects the number of ECR gates in the circuit.
}\vspace{.5cm}
\begin{tabular}{m{5cm}<{\centering}|m{3cm}<{\centering}|m{3cm}<{\centering}|m{3cm}<{\centering}}
\toprule [2pt]
Dilation method &
Sz.-Nagy &  
SVD & 
SVD + Walsh \\ 
\midrule [1pt]
Depth &
 148 &
 105 &
 60 \\
Number of 2-qubit gate & 28 &  22 &
 11 \\ 
\bottomrule [2pt]
\end{tabular}
\label{Tab:qc_depth_vs_Sz}
\vspace{3em}
\end{table}

\begin{figure}[!tbh]
\begin{center}
\includegraphics[width=1\linewidth]{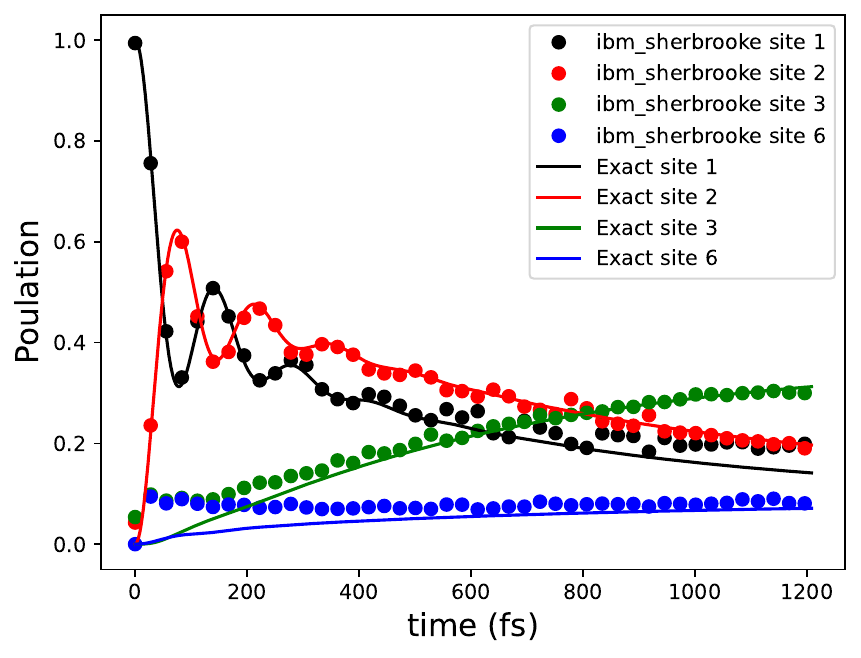}
\end{center}
\caption{Population dynamics of the FMO complex. Solid lines indicate the exact HEOM simulation results, while dotted scatter points represent quantum circuit results from the IBM Sherbrooke quantum computer. Each time point was sampled with 20000 shots. The projection subspace is defined as $S = \{11, 22, 33, 66\}$. Dynamic decoupling (XX sequence) and 2-qubit Clifford gate twirling error mitigation techniques were applied using Qiskit.
}
\label{fig:ibm3qb_fmo}
\end{figure}

Figure~\ref{fig:ibm3qb_fmo} shows the population dynamics of FMO obtained by running qHEOM on the IBM Sherbrooke quantum computer. The projection subspace $S = \{11, 22, 33, 66\}$ is consistent with Figure~\ref{fig:Aersimu}(b). Each quantum circuit was sampled 20000 times with error mitigation applied. Circuit complexity is uniform across time points (an example is provided in Table \ref{Tab:qc_depth}), and a sample circuit diagram is available in the SI.

Overall, the NISQ device accurately simulates the energy transfer dynamics in FMO, capturing both the coherent oscillations between site 1 and site 2 at short times and the growth of population at site 3 at longer times (the terminal site \cite{moix11}). Similar to the observations in Figure~\ref{fig:ibm3qb_CPC60}, slight deviations in NISQ results appear when $P(t)$ approaches zero, particularly in the initial populations of sites 3 and 6, and in the long-term population of site 1.

In both Figure~\ref{fig:ibm3qb_CPC60} and Figure~\ref{fig:ibm3qb_fmo}, the projected subspace $S$ includes four states, resulting in a nearly identical depth for the 3-qubit circuits in both cases. However, compared to the \ce{CPC60} results in Figure~\ref{fig:ibm3qb_CPC60}, the NISQ results for the FMO complex are more accurate. This is because, in the FMO case, the largest singular value $\sigma_0$ of $G(t)$ in eq~\ref{Eq:singval_Gt} is less than 1, whereas, for \ce{CPC60}, $\sigma_0$ exceeds 1. 
For the same $P_{exact}$ in eq~\ref{Eq:measure_error}, we have $N_{exact} = N_c (P_{exact}/\sigma_0)^2$. A smaller $\sigma_0$ thus results in a larger $N_{exact}$, reducing the deviation of the error factor $(\sqrt{1 \pm N_{err}/N_{exact}})$ from 1.

\begin{figure}[!tbh]
\begin{center}
\includegraphics[width=1\linewidth]{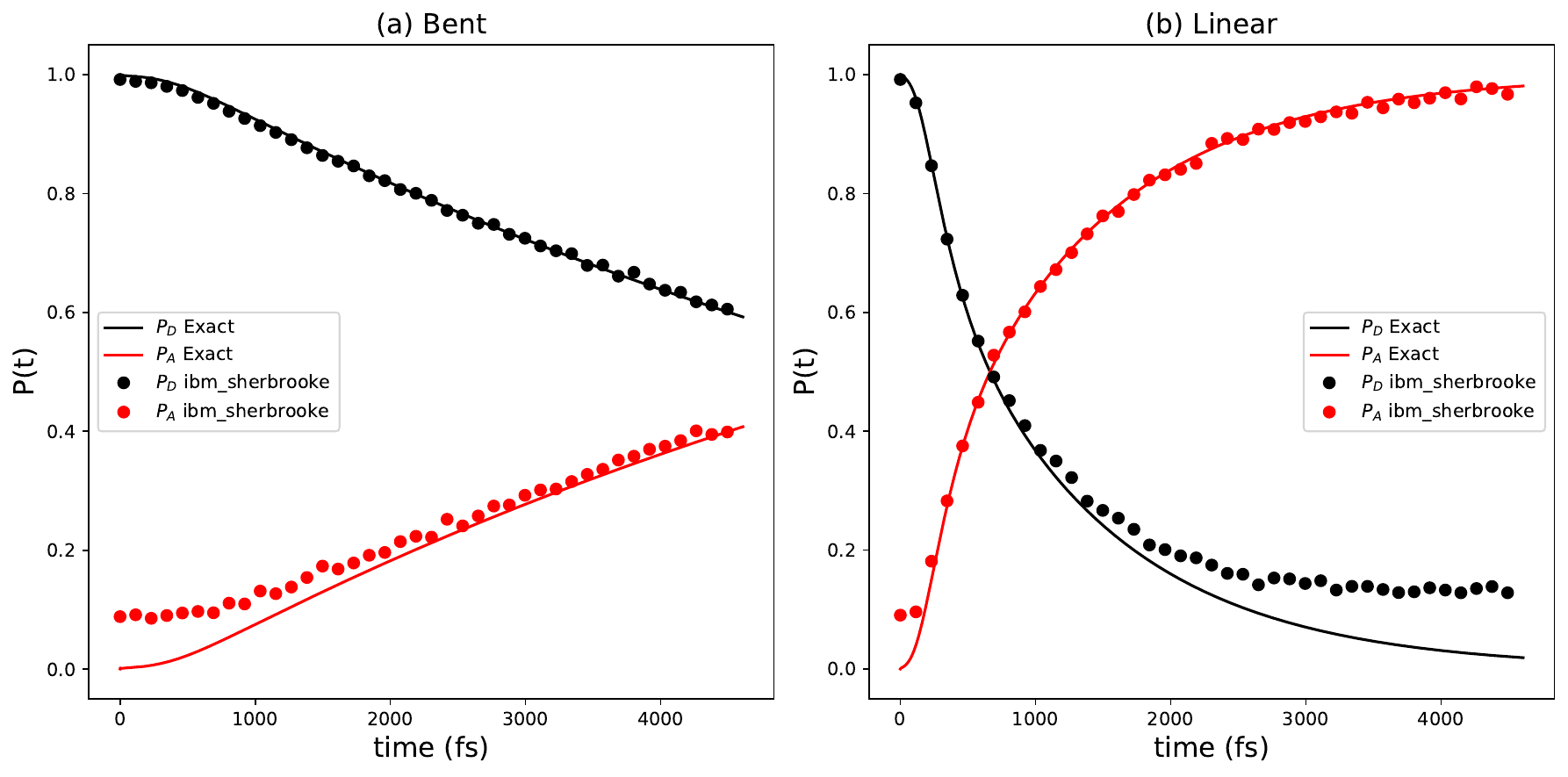}
\end{center}
\caption{Population dynamics of the \ce{CPC60} molecular triad. The projection subspace is defined as $S = \{DD, AA\}$, yielding 2-qubit quantum circuits. Solid lines denote the exact HEOM simulation results, while scatter points represent quantum circuit results from the IBM Sherbrooke quantum computer. Each time point was sampled with 20000 shots, with error mitigation applied.
}
\label{fig:ibm2qb_sbm}
\end{figure}

The quantum algorithm based on the dilation of $G(t)$ used in this work has the advantage of allowing the selection of projection subspace towards reducing the circuit depth, thereby lowering the effect of noise.\cite{wang23}
Figure~\ref{fig:ibm2qb_sbm} shows the population dynamics of the molecular triad obtained using the IBM Sherbrooke quantum computer, where the projected subspace is restricted to $S=\{DD,AA\}$. 
In this case, the corresponding dilation circuit involves only 2 qubits, with examples of the circuit shown in the SI. 
As shown in the figure, compared to Figure~\ref{fig:ibm3qb_CPC60}, the accuracy of the NISQ results has improved significantly. 
For both conformations, the NISQ results closely match the exact results, with only minor deviations when the exact population approaches zero. The reasons for these errors are the same as discussed earlier in Figure~\ref{fig:ibm3qb_CPC60} and Figure~\ref{fig:ibm3qb_fmo}.
The improved accuracy of the NISQ results is attributed to the significant reduction in circuit depth: in Table \ref{Tab:qc_depth}, due to the decrease in the number of qubits, the example circuit depth is reduced from 60 to 15, and the number of 2-qubit gates decreases from 11 to 2.

\begin{table}
\begin{tabular}{m{4cm}<{\centering}|m{3cm}<{\centering}|m{5cm}<{\centering}}
\toprule [2pt]
 & Depth &
Number of 2-qubit gates  \\ 
\midrule [1pt]
3 qubits \ce{CPC60} & 60 &  11\\
3 qubits FMO & 67 &  12 \\ 
2 qubits \ce{CPC60} & 15 &  2\\
2 qubits FMO & 17 &  2\\
\bottomrule [2pt]
\end{tabular}
\caption{Circuit complexities for different examples from Figure \ref{fig:ibm3qb_CPC60} to Figure \ref{fig:ibm2qb_fmo}. As specific cases, the circuit for \ce{CPC60} uses the propagator for the linear conformation at $t=2073.5$~fs, while for FMO uses the propagator at $t=612.0$~fs.
The 3-qubit \ce{CPC60} and 3-qubit FMO correspond to projection subspaces $S=\{DD,DA,AD,AA\}$ and $S=\{11,22,33,66\}$, respectively, while the 2-qubit \ce{CPC60} and 2-qubit FMO correspond to 
$S=\{DD,AA\}$ and $S=\{11,22\}$, respectively.
The circuit compilation settings are the same as those in Table~\ref{Tab:qc_depth_vs_Sz}.}
\label{Tab:qc_depth}
\vspace{2em}
\end{table}

The same approach can be applied to the FMO complex. Note that, as shown below eq~\ref{Eq:MPPsit}, the constraint $\mathcal{P}|\Psi(t=0)\rangle = |\Psi(t=0)\rangle$ imposed on the projection operator means that the projection subspace must include the initial site $|1\rangle$.
For the 2-qubit circuit, $S$ should take the form of $\{11, ij\}$. 
Therefore, to measure the populations of sites 1, 2, 3, and 6 at a given time, we need three 2-qubit circuits corresponding to the projection subspaces $S_1=\{11,22\}$, $S_2=\{11,33\}$, and $S_3=\{11,66\}$. 
The decomposition of the projection subspace can significantly reduce the circuit depth. 
As an example shown in Table \ref{Tab:qc_depth}, 
after decomposing the $S=\{11,22,33,66\}$ into $S_1$, $S_2$, and $S_3$, the circuit with a depth of 67 and 12 two-qubit gates is transformed into three circuits, each with a depth of around 17 and containing two two-qubit gates.

\begin{figure}[!ht]
\begin{center}
\includegraphics[width=1\linewidth]{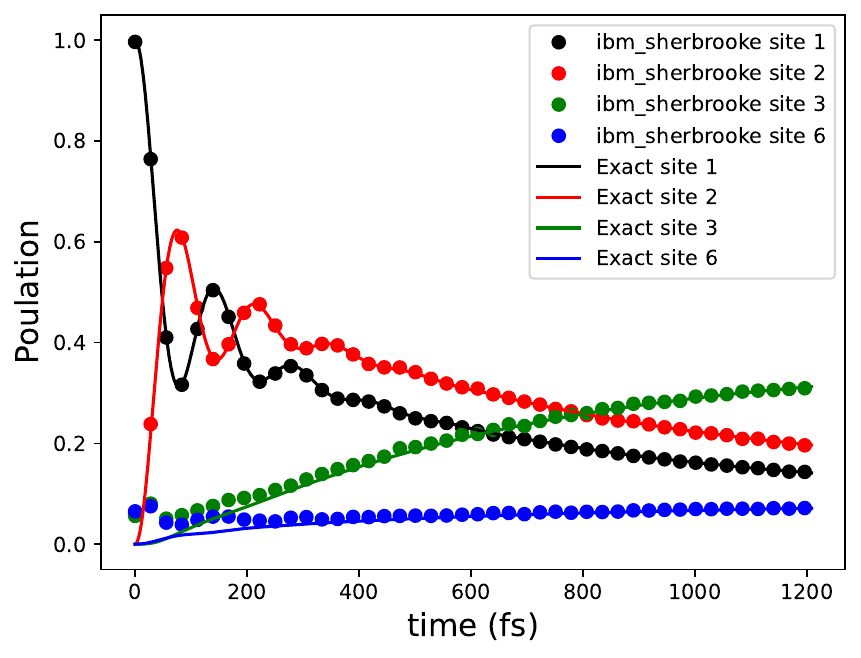}
\end{center}
\caption{Population dynamics of the FMO complex. 
The solid lines represent the exact HEOM results, and the dot scatter points represent the quantum circuit results from the IBM Sherbrooke quantum computer. 
To calculate the populations of states $|1\rangle$, $|2\rangle$, $|3\rangle$, and $|6\rangle$ using two-qubit circuits, three different projection subspaces are employed: $S_1=\{11,22\}$, $S_2=\{11,33\}$, and $S_3=\{11,66\}$.
Each time point is measured by $20000$ shots with error mitigation.}
\label{fig:ibm2qb_fmo}
\end{figure}

The reduction in circuit complexity has significantly improved the accuracy of the NISQ results. 
Figure \ref{fig:ibm2qb_fmo} shows the IBM Sherbrooke results for the FMO complex, where we use $S_1=\{11,22\}$, $S_2=\{11,33\}$, and $S_3=\{11,66\}$ to construct 2-qubit quantum circuits. 
Except for slight deviations in the population of sites 3 and 6 at early time points $t<200$~fs, the IBM Sherbrooke results align almost perfectly with the exact results. This demonstrates that our quantum algorithm can achieve high precision in simulating the population dynamics of the FMO complex on actual NISQ devices.

\pagebreak
\section{Concluding Remarks}\label{Sec: conclusion}

We have introduced the qHEOM quantum algorithm and demonstrated its accuracy and practical utility by using it to simulate charge transfer in a solvated molecular triad and excitation energy transfer in the FMO complex, on NISQ devices.
We have shown that the dynamics in both cases are non-Markovian and beyond the weak system-bath coupling limit, thereby rendering a description via the Lindblad quantum master equation inadequate. 
Specifically, we found that the TCL-Redfield equation with the RWA (quantum master equation in the Lindblad form) is limited to the coherent dynamics regime, fails to capture the charge transfer rate in the molecular triad, and provides only qualitative results for energy transfer in FMO. 

qHEOM implements the SVD dilation to convert the non-unitary HEOM propagator into unitary gates and utilizes the Walsh operator representation to efficiently represent the diagonal unitary operator within the dilation circuit. In doing so, qHEOM significantly reduces circuit complexity compared to algorithms based on Sz.-Nagy dilation.\cite{wang23,lyu24}  Further reduction in circuit depth and complexity is achieved by projecting the HEOM propagator onto the subspace of quantities of interest.  Since different subspaces are independent, this approach also allows for parallel quantum computing implementations on multiple circuits. 

Our qHEOM simulations performed on the IBM NISQ device show that device noise can lead to significant discrepancies. The discrepancy is found to be related to the properties of the propagator.  When the largest singular value of the propagator is relatively large, it can amplify this discrepancy. Consequently, for circuits of similar complexity, the NISQ results for energy transfer in the FMO complex are more accurate than those for charge transfer in the molecular triad.

Reducing the size of the projection subspace can mitigate device-induced errors. By limiting the projection subspace to include only two states, the IBM device achieved highly accurate quantitative results. Notably, for energy transfer in the FMO complex, the NISQ results are almost perfectly aligned with numerically exact benchmark results. 

When simulating open quantum system dynamics via qHEOM on IBM quantum computers, error mitigation techniques included in Qiskit can effectively reduce errors caused by device noise. 
However, the noise susceptibility of qubits on NISQ devices still limits the problems these machines can address. Consequently, this work employs small projected subspaces to obtain meaningful results. Nevertheless, the proposed algorithm is not inherently restricted to such subspaces. If deep quantum circuits involving a large number of qubits can be executed with lower error rates, the exact propagation of HEOM, a significant challenge for classical computers, could be achieved without using projected subspaces by constructing the propagator for the full Hilbert space. 

To this end, future efforts could explore different dynamical decoupling sequences \cite{khodjasteh05,uhrig07,pokharel18,ezzell23} to suppress errors. 
Building on this foundation, advancements in fault-tolerant quantum computing (FTQC) \cite{campbell17,katabarwa24,ian22,andreasson19} may provide a pathway to integrate the proposed algorithm with quantum error correction. In this scenario, the physical qubits in our circuits would be replaced by logical qubits, encoded using quantum error correction codes (QECC) \cite{shor95,steane97,devitt13,gottesman10}, to correct errors in qubits and further enhance computational reliability. 

We note that the quantum algorithm implemented in this work is quite general. Although we used it to implement the HEOM propagator, the same algorithm could be applied to implement propagators from other quantum master equations. Those propagators could be obtained using various numerical methods, such as TT-TFD,\cite{gelin17,borrelli21b,lyu23} path integral,\cite{topaler93,makri95,weiss08b,segal10,weiss13,makri20,bose22,walters24,seneviratne24} time-evolving matrix product operator (TEMPO),\cite{strathearn18,jorgensen19,richter22} GQME,\cite{zwanzig61a,shi03g,shi04a,zhang06,kelly16,montoya16,montoya17,pfalzgraff19,mulvihill19,mulvihill21b,mulvihill22,ng21,dan22,lyu23} and others. 
Once the propagator for the evolution of the system is obtained, the SVD dilation and Walsh operator representation implemented in this work can be applied to construct the corresponding quantum circuit. 

In future studies, we will use bosonic quantum devices to simulate the dissipative dynamics of chemical systems according to the quantum algorithm developed in this work.  Hybrid qubit-qumode quantum devices could offer significant advantages over traditional qubit-based quantum platforms,\cite{cabral24,dutta24,crane24,liu24,lyu24b} particularly for implementations of the HEOM methodology where the degrees of freedom of the environment are decomposed into several effective bosonic modes. 
Moreover, by selecting appropriate projection operators, the method introduced in this study can be extended to more complex models and interactions onto quantum circuits, such as the Holstein model,\cite{yan18,cainelli21,li24b} the Holstein-Hubbard model,\cite{nakamura21} and model systems with conical intersections.\cite{wang24}

\section{Acknowledgments}
Xiaohan Dan acknowledges Professor Qiang Shi, Professor Chen Wang, Ningyi Lyu, Delmar G. A. Cabral, Brandon Allen, and Yuanjun Shi for useful discussions and assistance with illustrations.
VSB acknowledges support from the National Science Foundation Engines Development Award: Advancing Quantum Technologies (CT) under Award Number 2302908, and partial support from the National Science Foundation Center for Quantum Dynamics on Modular Quantum Devices (CQD-MQD) under Award Number 2124511. E.G. acknowledges additional support from the NSF Grant No. CHE 2154114. VSB acknowledges a generous allocation of HPC time from NERSC and from the Yale Center for Research Computing.

\begin{suppinfo}
Includes a detailed description of the quantum circuits and additional figures.
\end{suppinfo}

\section{Code availability}
The Python code for qHEOM simulations is available at {\href{https://github.com/XiaohanDan97/qHEOM}{https://github.com/XiaohanDan97/qHEOM}}. 

\pagebreak
\bibliography{./quantum}

\clearpage
\renewcommand{\thefigure}{S\arabic{figure}}
\renewcommand{\thetable}{S\arabic{table}}
\setcounter{figure}{0}
\setcounter{table}{0}

\begin{center}
\bf{
{\large Supporting Information: Simulating Non-Markovian Quantum Dynamics on NISQ Computers Using the Hierarchical Equations of Motion}}
\end{center}


\maketitle

\section*{Quantum circuits}
In this section, we give examples of the quantum circuits used in the main text. 
All quantum circuits were compiled according to the topology of the corresponding IBM quantum computer, specifically the IBM Sherbrooke, with the basis gate set consisting of $X$, $\sqrt{X}$, $R_z$, and ECR gates. 
The compilation processes were executed using the Qiskit package.\cite{qiskit24}

\begin{figure}[H]
\begin{center}
\includegraphics[width=1\linewidth]{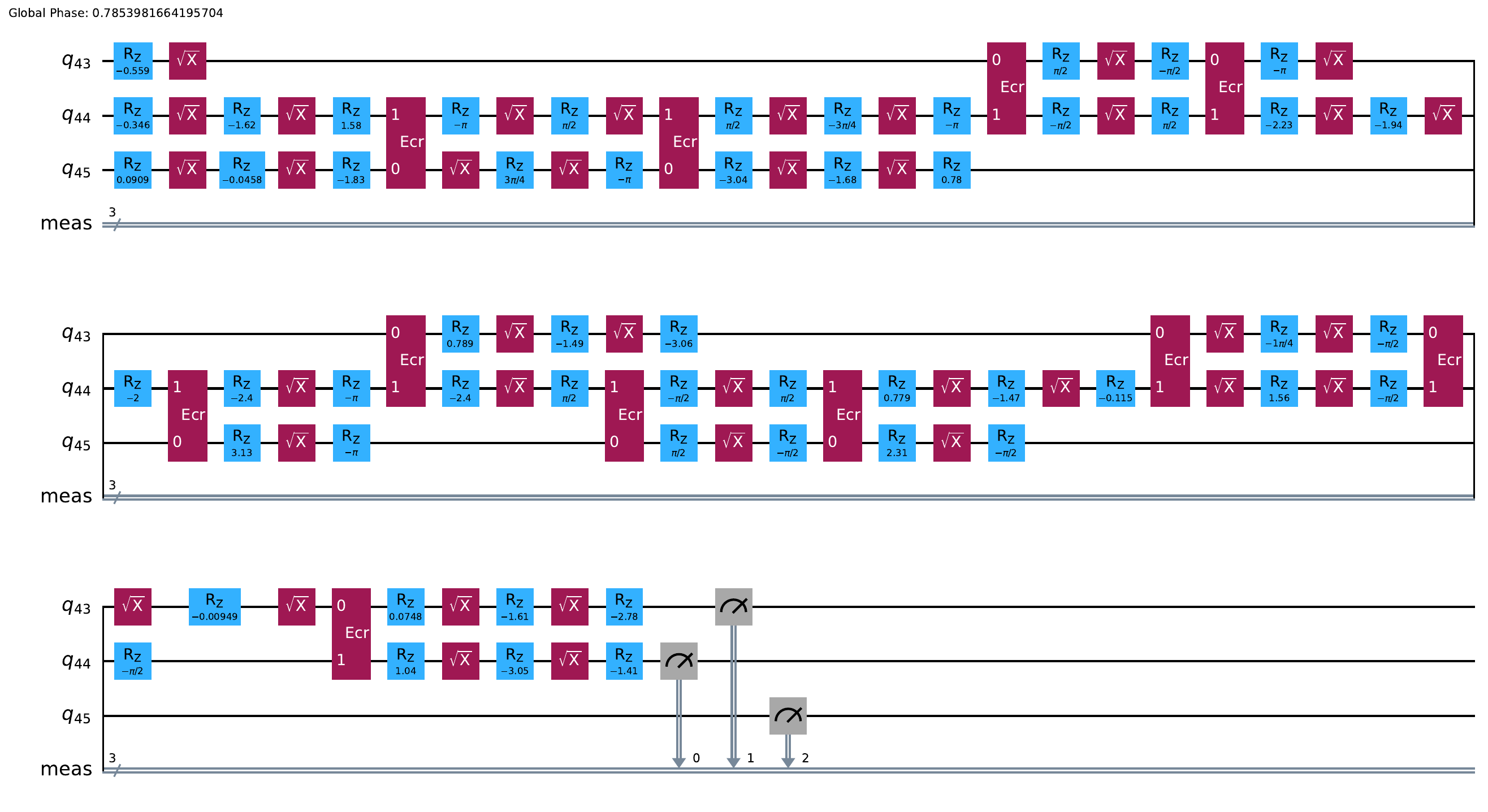}
\end{center}
\caption{Quantum circuit for the molecular triad charge transfer model in the linear conformation,
with the propagator at $t=2073.5$~fs.
The projection subspace $S=\{DD,DA,AD,AA\}$ corresponds to 3 qubits after dilation.
The circuit depth is 60, with the following gate counts: 55 $R_z$ gates, 38 $\sqrt{X}$ gates, and 11 ECR gates.}
\label{fig:qc_linear_3qubitSBM}
\end{figure}

\begin{figure}[H]
\begin{center}
\includegraphics[width=1\linewidth]{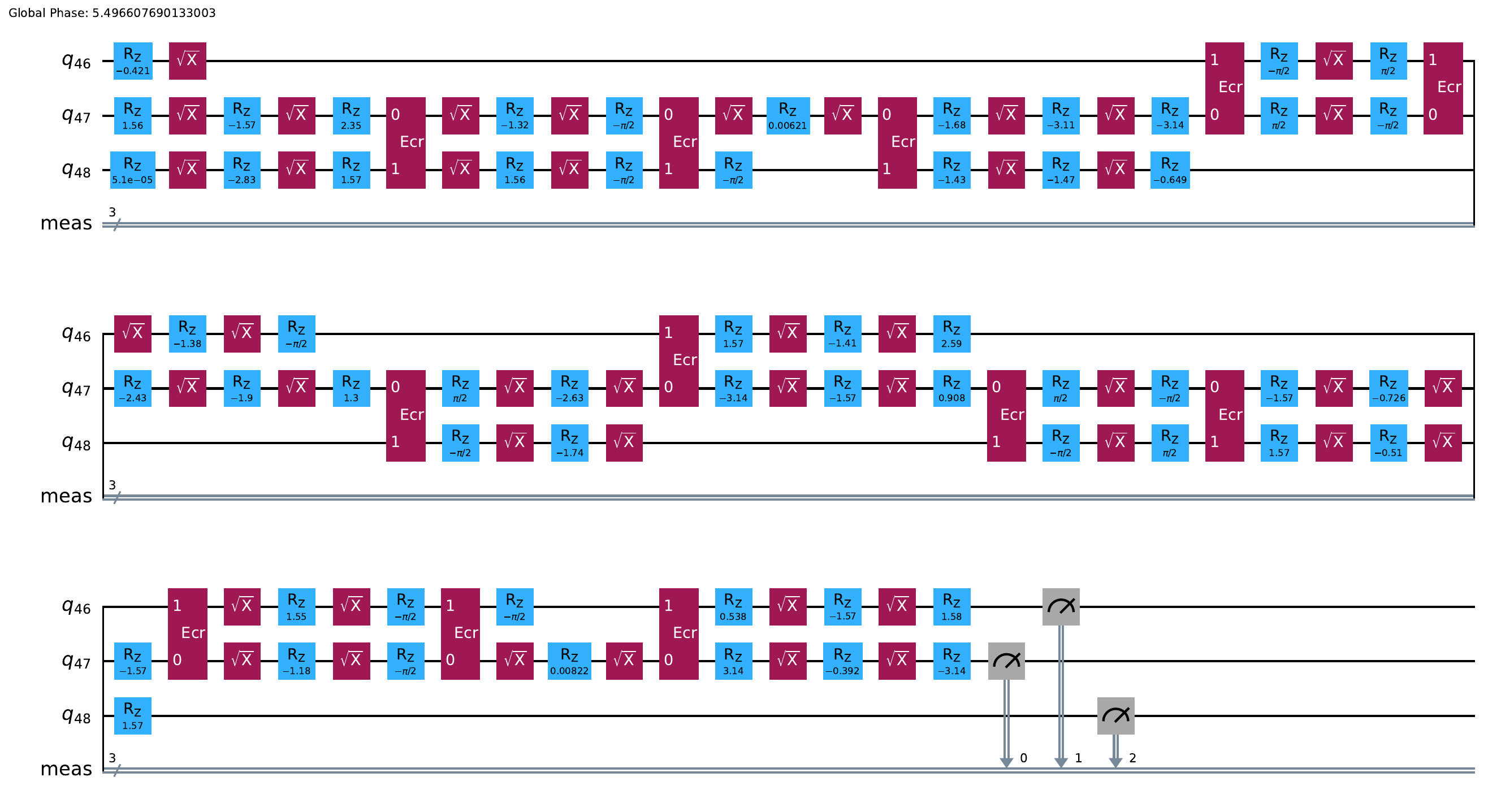}
\end{center}
\caption{3-qubit quantum circuit corresponding to the propagator at $t=612.0$~fs for the excitation energy transfer model in the FMO complex, the projection subspace $S=\{11,22,33,66\}$.  
The circuit depth is 67, with 60 $R_z$ gates, 45 $\sqrt{X}$ gates, and 12 ECR gates.}
\label{fig:qc_3qubitFMO}
\end{figure}

\begin{figure}[H]
\begin{center}
\includegraphics[width=1\linewidth]{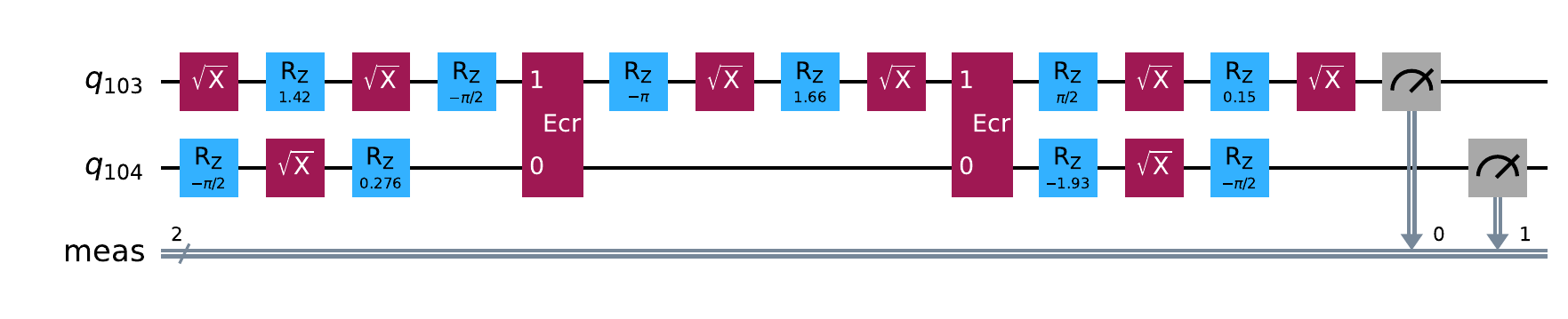}
\end{center}
\caption{Quantum circuit for the 2-qubit linear conformation molecular triad charge transfer model (the projection subspace $S=\{DD,AA\}$), corresponding to the propagator at $t=2073.5$~fs.
The circuit depth is 15, with the following gate counts: 10 $R_z$ gates, 8 $\sqrt{X}$ gates, and 2 ECR gates.}
\label{fig:qc_linear_2qubitSBM}
\end{figure}

\begin{figure}[H]
\begin{center}
\includegraphics[width=1\linewidth]{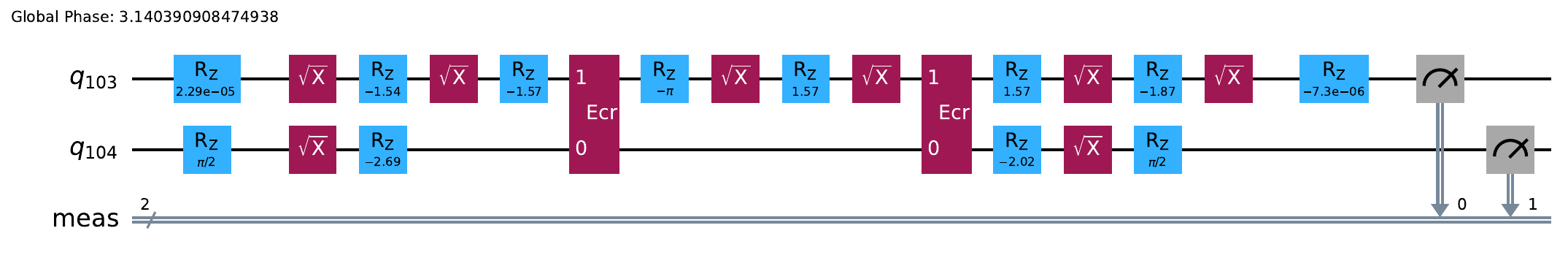}
\end{center}
\caption{2-qubit quantum circuit corresponding to the propagator at $t=612.0$~fs for the excitation energy transfer model in the FMO complex, with the projection subspace $S=\{11,22\}$. The circuit depth is 17, with 12 $R_z$ gates, 8 $\sqrt{X}$ gates, and 2 ECR gates.
}
\label{fig:qc_2qubitFMO}
\end{figure}

\section*{Results without error mitigation}
This section presents the NISQ results without error mitigation techniques for charge transfer in the molecular triad and energy transfer in the FMO complex.

\begin{figure}[!tbh]
\begin{center}
\includegraphics[width=1\linewidth]{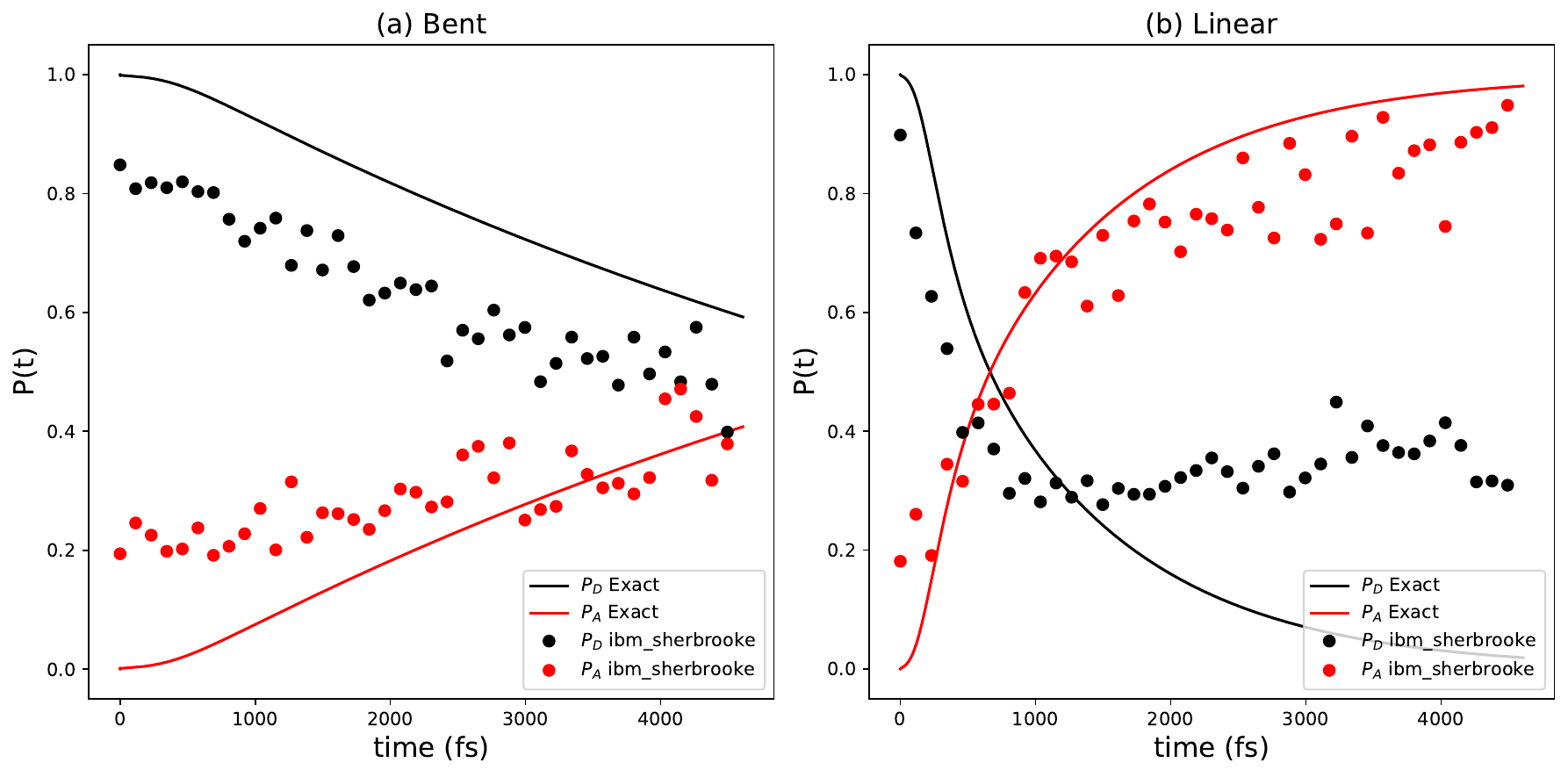}
\end{center}
\caption{Population dynamics of the molecular triad without error mitigation: (a) Bent conformation, (b) linear conformation. $P_D$ denotes the population in the $\pi\pi^*$ (donor) state, while $P_A$ denotes 
the population in the ${\rm CT1}$ (acceptor) state. 
The solid lines represent the exact HEOM results, and the scatter points represent the quantum circuit results from the IBM Sherbrooke computer. 
The projection subspace $S=\{DD,DA,AD,AA\}$. 
Each time point is measured at $20000$ shots.}
\label{fig:ibm3qb_CPC60SI}
\end{figure}

\begin{figure}[H]
\begin{center}
\includegraphics[width=1\linewidth]{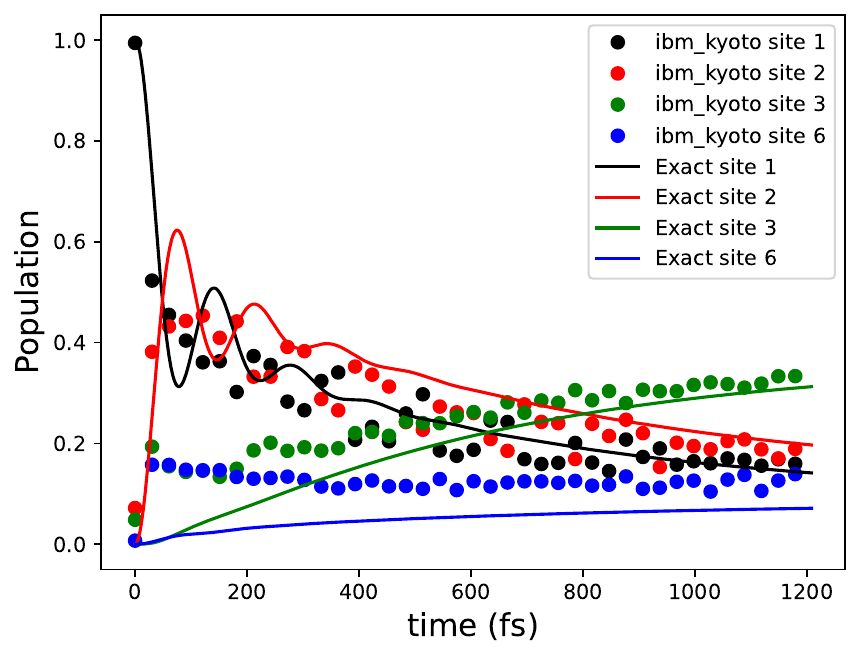}
\end{center}
\caption{Population dynamics of the FMO complex without error mitigation. The solid lines represent the exact HEOM simulation results, and the scatter points represent the quantum circuit results from the IBM Kyoto quantum computer. 
The projection subspace $S=\{11,22,33,66\}$.
Each time point is measured at $20000$ shots.}
\label{fig:ibm_FMO}
\end{figure}



\end{document}